\documentclass[useAMS,usenatbib]{mn2e}
\usepackage{mathptmx}
\usepackage[british]{babel}
\usepackage{ifthen}
\usepackage[breaklinks]{hyperref}
\usepackage{etoolbox}
\usepackage{xcolor}
\usepackage[pdftex]{graphicx} 
\usepackage{microtype}
\usepackage{xspace}
\usepackage{amssymb,amsmath}
\usepackage{paralist}
\usepackage{placeins}
\usepackage{float}
\usepackage{enumitem}
\usepackage{multirow}
\usepackage{bm}

\newcommand{\ie}{{\it i.e.}\xspace}
\newcommand{\eg}{{\it e.g.}\xspace}

\newcommand{\etal}{{\it et al.}\xspace}
\newcommand{\kpc}{\ensuremath{\,{\rm kpc}}\xspace}
\newcommand{\pc}{\ensuremath{\,{\rm pc}}\xspace}
\newcommand{\Myr}{\ensuremath{\,{\rm Myr}}\xspace}
\newcommand{\Gyr}{\ensuremath{\,{\rm Gyr}}\xspace}
\newcommand{\kms}{\ensuremath{\,{\rm km}\,{\rm s}^{-1}}\xspace}
\newcommand{\mags}{\ensuremath{\,{\rm mag}}\xspace}

\newcommand{\dg}{\ensuremath{^{\circ}}\xspace}

\newcommand{\K}{\ensuremath{{K_s}}\xspace}
\newcommand{\Kband}{\K-band\xspace}

\newcommand{\murc}{\ensuremath{{\mu_{{\rm RC}}}}\xspace}
\newcommand{\msun}{\ensuremath{{M_\odot}}\xspace}

\newcommand{\mone}{\ensuremath{{3.6\micron}}\xspace}
\newcommand{\Mtwo}{\ensuremath{{[4.5\mu]}}\xspace}
\newcommand{\mtwo}{\ensuremath{{4.5\micron}}\xspace}

\newcommand{\HK}{{\ensuremath{C}\xspace}}
\newcommand{\MHK}{{\ensuremath{M_C}\xspace}}
\newcommand{\EHK}{{\ensuremath{E_C}\xspace}}
\newcommand{\extlaw}{{\ensuremath{R_K}\xspace}}
\newcommand{\eqn}[1]{{equation (\ref{eq:#1})\xspace}}

\newcommand{\fig}[1]{{Fig. \ref{fig:#1}\xspace}}

\newcommand{\sgra}{Sgr A*}	

\AtBeginDocument{

\def\equationautorefname~#1\null{equation~(#1)\null}
}

\title[The Milky Way's Bar Outside the Bulge]{The Structure of the Milky Way's Bar Outside the Bulge}
 
\author[Wegg \etal]
{Christopher Wegg$^{1}$\thanks{E-mail: wegg@mpe.mpg.de}, Ortwin Gerhard$^{1}$ and Matthieu Portail$^{1}$\\
$^1$Max-Planck-Institut f\"ur Extraterrestrische Physik, Giessenbachstrasse, 85748 Garching, Germany.}
\pagerange{\pageref{firstpage}--\pageref{lastpage}} \pubyear{2015}
 
\begin{document}
\label{firstpage}
\maketitle

\begin{abstract}
While it is incontrovertible that the inner Galaxy contains a bar, its
structure near the Galactic plane has remained uncertain, where
extinction from intervening dust is greatest. We investigate here the
Galactic bar outside the bulge, the long bar, using red clump giant
(RCG) stars from UKIDSS, 2MASS, VVV, and GLIMPSE. We match and combine
these surveys to investigate a wide area in latitude and longitude, 
$ |b| \le 9\dg$ and $ |l| \le 40\dg$.  We find: 
\begin{inparaenum}
\item The bar extends to $l \sim 25 \dg$ at $|b| \sim 5\dg$ from the
Galactic plane, and to $l \sim  30 \dg$ at lower latitudes.  
\item The long bar
has an angle to the line-of-sight in the range $(28-33) \dg$, consistent with studies of the
bulge at $|l|<10\dg$. 
\item The scale-height of RCG stars smoothly
transitions from the bulge to the thinner long bar. 
\item There is
evidence for two scale heights in the long bar. We find a $\sim 180 \pc$ thin
bar component reminiscent of the old thin disk near the sun, and a
$\sim 45 \pc$ super-thin bar component which exists predominantly towards the
bar end. 
\item Constructing parametric models for the RC magnitude
distributions, we find a bar half length of $5.0\pm 0.2 \kpc$ for the
2-component bar, and $4.6 \pm 0.3 \kpc$ for the thin bar component alone. 
\end{inparaenum}
We conclude that the Milky Way contains a central box/peanut bulge which
is the vertical extension of a longer, flatter bar, similar as seen in
both external galaxies and N-body models.
\end{abstract}
\begin{keywords}
Galaxy: bulge -- Galaxy: center -- Galaxy: structure.
\end{keywords}


\begin{figure*}
\includegraphics[width=\linewidth]{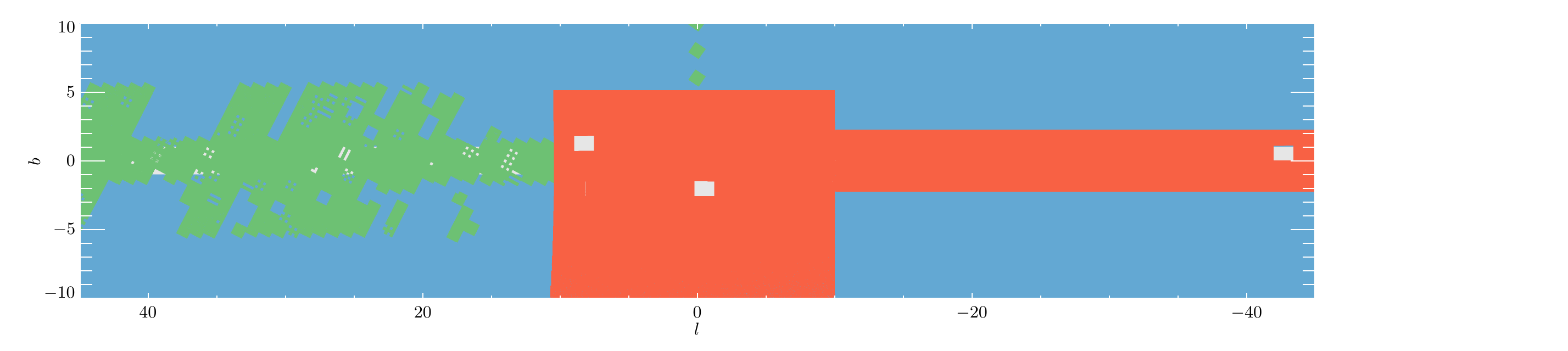}
\includegraphics[width=\linewidth]{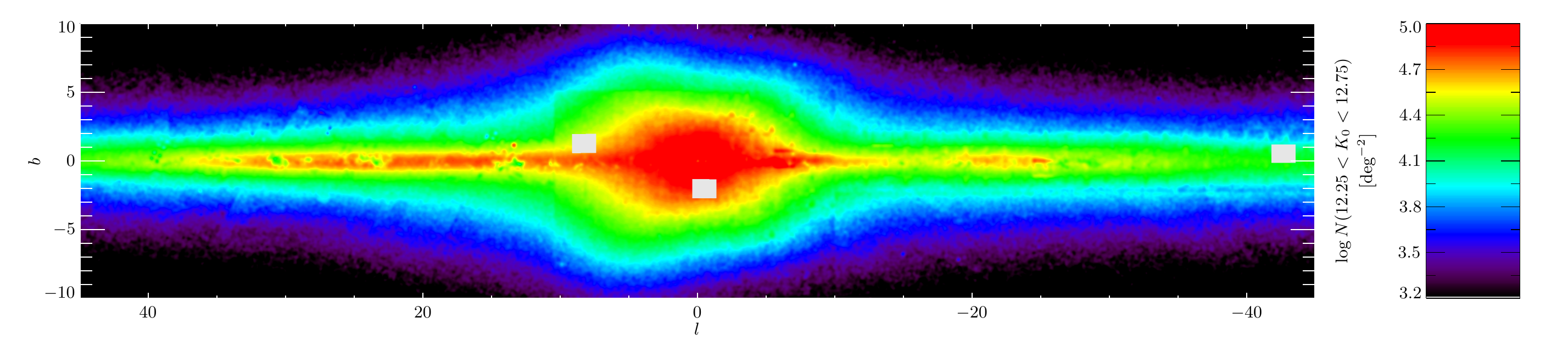}
\caption{In the top figure we show the surveys used in this study. We use, in order of preference, VVV in red, UKIDSS in green, and 2MASS in blue. Grey regions are those without data of sufficient depth \ie close to the plane without VVV or UKIDSS data where 2MASS is insufficient. In the lower figure we show the surface density of stars in the \Kband in the extinction-corrected magnitude range $12.25 < K_0 < 12.75$. Asymmetric number counts in $l$ close to the plane are a result of non-axisymmetry due to the long bar. The star counts are smoothed with a Gaussian kernel of $\sigma=0.1\dg$. Extinction is corrected using the $H-K$ colour excess as in  \autoref{eq:mujhk} (\ie $K_0 \equiv \mu_K + M_{K,{\rm RC}}$) and data outside the colour bar range are plotted at its limit. \label{fig:extcormap}}
\end{figure*}


\section{Introduction}

Gas kinematics and near-Infrared (NIR) photometry  \citep[\eg][]{Binney:91,Weiland:94} have shown that the Galactic bulge at longitudes $|l|<10\dg$ contains a bar-like structure. These results have since been confirmed, and the barred bulge has been characterised and mapped  with increasing accuracy. Using star counts of red clump giants (RCGs), \citet{Stanek:94,Stanek:97} reached a similar conclusion that the bulge was barred, a result since confirmed with increasing confidence and accuracy \citep{Babusiaux:05,Rattenbury:07}. The studies using RCGs agree with results from brighter giant star counts \citep{LopezCorredoira:05} and the COBE integrated NIR emission \citep{Dwek:95,Binney:97,Freudenreich:98,Bissantz:02}. Most authors have found a bar angle, defined as the angle of the major axis of the bar to the sun line-of-sight, in the range $20-30\dg$. Recent interest was stimulated by the discovery that close to the bulge minor axis at $|b|\gtrsim5\dg$ the RCGs have two distinct magnitudes \citep{McWilliam:10,Nataf:10}. This is because these lines-of-sight pass through both arms of an X-shaped structure which is characteristic of boxy/peanut (B/P) bulges in barred galaxies \citep{McWilliam:10,Ness:12}. These B/P bulges arise naturally in N-body simulations of disk galaxies \cite[\eg][]{Athanassoula:05,Inma:06} and are common in external galaxies \citep{Bureau:06,Laurikainen:14}. This has stimulated further investigation of the full three dimensional structure of the bulge beginning with \citet{Saito:11} using 2MASS data. Vista Variables in the Via Lactea (VVV) has provided significantly deeper and more complete data and this was recently used to map the B/P bulge  non-parametrically in three dimensions by \citet{Wegg:13}. 

It has also been suggested in NIR star counts using in-plane data at $10\dg<l<30\dg$ that there is a less vertically extended $\approx 4\kpc$ bar at $\approx 45^\circ$ to the line of sight \citep{Hammersley:94}. This component has been termed the long bar. The existence of the long bar was confirmed using RCGs by \citet{Hammersley:00} and subsequently with increasingly more powerful NIR \citep{LopezCorredoira:06,CabreraLavers:07,CabreraLavers:08} and longer wavelength GLIMPSE data \citep{Benjamin:05,Zasowski:thesis}.

Understanding the nature of the long bar and its structure, amplitude, length and pattern speed is of great importance for many areas of Milky Way study. It influences for example the disk outside the bar \citep{Minchev:10}, the kinematics in the solar neighbourhood \citep{Dehnen:00}, and the observed non-circular gas flow \citep{Bissantz:03}.

One important unresolved issue is the relationship between the bar in the Galactic bulge at $|l|<10\dg$, and the long bar outside it. Studies of the long bar at $l\gtrsim 10\dg$ have often found a larger bar angle \citep[typically $\approx 45\dg$, beginning with][]{Hammersley:94} in comparison to the three-dimensional bulge \citep[typically $\approx 25-30\dg$ \eg $(27\pm2)\dg$: ][]{Wegg:13}. It has therefore been suggested that the Galaxy contains two bars, with the central three-dimensional bar not aligned to the long bar. However an in-plane long bar misaligned with a triaxial three-dimensional bar is difficult to reconcile dynamically; the suggested length ratio is low and therefore the mutual torques are strong. It has instead been suggested that rather than two distinct bars, the long bar is the in-plane extension of the central three-dimensional boxy/peanut structure structure \citep{MartinezValpuesta:11,Romero:11}. One of the motivations for this study is to help resolve this controversy. 

Throughout we use the terminology that the bar outside the bulge region at $|l|>10\dg$ is the long bar, regardless of the details of thickness, bar angle, or alignment with the barred bulge.  

Our primary indicator of bar structure are RCGs which are core helium burning stars and provide an approximate standard candle \citep{Stanek:94}. We combine several surveys to have the widest view and the greatest possible scope on the bar density distribution. In the \Kband we use data from
\begin{inparaenum}
\item the United Kingdom Infrared Deep Sky Survey (UKIDSS) Galactic Plane Survey \citep[GPS,][]{Lucas:08}, 
\item the VVV survey \citep{Saito:12} and,
\item to extend the study further from the galactic plane than previous studies, we augment this with 2MASS data \citep{Skrutskie:06}.
\end{inparaenum}
We homogenise the analysis of the surveys using a common photometric system and identify RCGs statistically in magnitude distributions rather than in colour-magnitude diagrams since this has worked well in the bulge \citep[\eg][]{Nataf:13,Wegg:13}.

We verify our results where possible using data at 3.6\micron~and 4.5\micron, which is significantly less affected by extinction, taken from the  Galactic Legacy Mid-Plane Survey Extraordinaire (GLIMPSE) survey on the Spitzer space telescope \citep{Benjamin:05}. Because this data only covers $|b| \lesssim 1\dg$ we use the \Kband as our primary data, but the GLIMPSE data remains very important for cross checks, particularly of dust extinction. 

This work is organised as follows: 
In \autoref{sec:magdists} we describe the data and construction of magnitude distributions for the stars in bulge and bar fields.
In \autoref{sec:rcfits} we fit the red clump stars in these magnitude distributions and discuss the features of these fits.
In \autoref{sec:slicefits} we examine the vertical structure of the fitted red clump stars in longitude slices,
and in \autoref{sec:fitrho} we derive densities which fit and best fit the observed magnitude distributions.
We discuss our results and place them in context in \autoref{sec:discuss}, 
and finally conclude in \autoref{sec:conclude}.

\section{Magnitude Distribution Construction}
\label{sec:magdists}

A number of steps are required to combine the surveys (UKIDSS, VVV, 2MASS, GLIMPSE) and bands ($H$, \K, \mone, \mtwo) and construct consistent magnitude distributions: The surveys must be transformed to the same photometric system, extinction corrected, and to compare bands and convert to distances we require the characteristic magnitudes and colours of RCGs.

\subsection{$H$ and \Kband data}

The first step in construction of the magnitude distributions is to transform all the surveys to the same photometric system. We choose to convert the UKIDSS and VVV surveys to the 2MASS system using the methods and transformations described in appendix \ref{sec:xforms}.

\subsubsection{Extinction Correction}
\label{sec:kextcor}

Extinction is then corrected for on a star-by-star basis assuming that all stars are red-clump giants (RCGs). We primarily work with the RCG \Kband distance modulus, $\mu_K$, where we calculate the \Kband extinction from the $H-\K$ reddening:
\begin{equation}
\mu_K = \K - \overbrace{\frac{A_\K}{E(H-\K)} \underbrace{\left[ (H - \K) - (H-\K)_{\rm RC} \right]}_{\mbox{Reddening}}}^{\mbox{Extinction Correction}}- M_{\K,{\rm RC}} ~,\label{eq:mujhk}
\end{equation}
where $(H-\K)_{\rm RC}$ is the intrinsic $H-\K$ colour of RCGs, $M_{\K,{\rm RC}}$ is the absolute \Kband magnitude of RCGs, and $\frac{A_\K}{E(H-\K)}$ is a constant which depends on the extinction law.

The method of correcting for extinction on a star-by-star basis in \autoref{eq:mujhk} is similar to other studies with data close to the Galactic plane \citep[\eg][]{Babusiaux:05,CabreraLavers:08}. In the Galactic bulge most of the dust lies in a foreground screen making extinction correction using extinction maps possible. In-plane this is not true and necessitates the method of extinction correction in \autoref{eq:mujhk}.

In the near-infrared bands used in this work, the vast majority of the stars have colours close to that produced by a Rayleigh-Jeans spectrum, therefore this star-by-star extinction correction using the $H-\K$ is relatively accurate \citep[\eg the NICE method,][]{Lada:94}. Moreover, we are concerned with RCGs in this work, which as well as being good standard candles naturally have very similar colours. For easy comparison between bands we work with the distance modulus, under the assumption that the stars are RCGs. Other stars form a background which we do not study. We use $H-\K$ to correct rather than $J-\K$ because the $H$-band has lower extinction and therefore remains deeper in high extinction regions. In addition $H-\K$ is more constant across stellar types than $J-\K$ improving the extinction correction \citep{Majewski:11}.

To utilise \autoref{eq:mujhk} we require two additional ingredients: a value for $\frac{A_\K}{E(H-\K)}$, and the magnitude of RCGs for $(H-\K)_{\rm RC}$ and $M_{\K,{\rm RC}}$.

To estimate $\frac{A_\K}{E(H-\K)}$ we use the value $A_H/A_\K=1.73$ measured by \citet{Nishiyama:09} using red clump stars close to the galactic plane towards the galactic center. This corresponds to $\frac{A_\K}{E(H-\K)}=1.37$. In principle the $H$-band could be used as an additional confirmation of our results, however the adopted $A_H/A_\K$ results in a $H$-band equivalent to \autoref{eq:mujhk} for $\mu_H$ which is completely degenerate with $\mu_K$. Instead we use the GLIMPSE survey to confirm results where possible because the longer wavelengths are less susceptible to dust extinction.

We estimate the absolute magnitudes of red clump stars in a two step process. First we estimate colours of red clump stars using the Padova isochrones. Constructing the luminosity function from the solar metallicity 10\Gyr isochrone with a Kroupa IMF and fitting a gaussian to the red clump results in red clump colour of $H-\K=0.09$. This colour (and the GLIMPSE colours utilised later) are largely independent of age and metallicity, changing by less than 0.03\mags when varying age from 1\Gyr to 15\Gyr and metallicity from $[{\rm Fe/H}]=-0.7$ to $[{\rm Fe/H}]=0.17$. Finally we use the calibration $M_{\K,{\rm RC}}=-1.72$ which results in bulge red clump stars being approximately symmetric about $R_0\approx8.3\kpc$ \citep{Wegg:13}.      

\begin{figure}
\includegraphics[width=\linewidth]{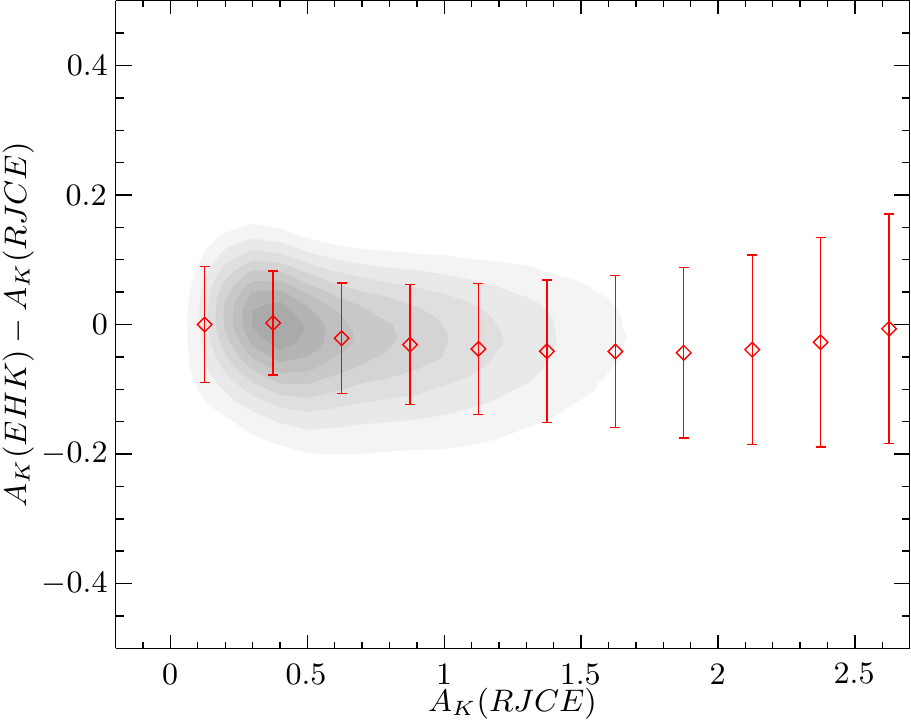}
\caption{The difference in \Kband extinction calculated using two methods. $A_\K(EHK)$ is calculated using the $H-\K$ reddening from \autoref{eq:mujhk}. $A_\K(RJCE)$ is the \Kband extinction calculated by converting the $H-\Mtwo$ reddening from \autoref{eq:muglimpse} to $A_\K$. All stars with $12.5<\mu_K<14$ are included. Contours are equally spaced in density at 10, 20... 90\% of the peak density. Error bars represent the sigma clipped mean and standard deviation binned as a function of $A_\K(RJCE)$. \label{fig:ext_ext}}
\end{figure}

\begin{figure}
\includegraphics[width=\linewidth]{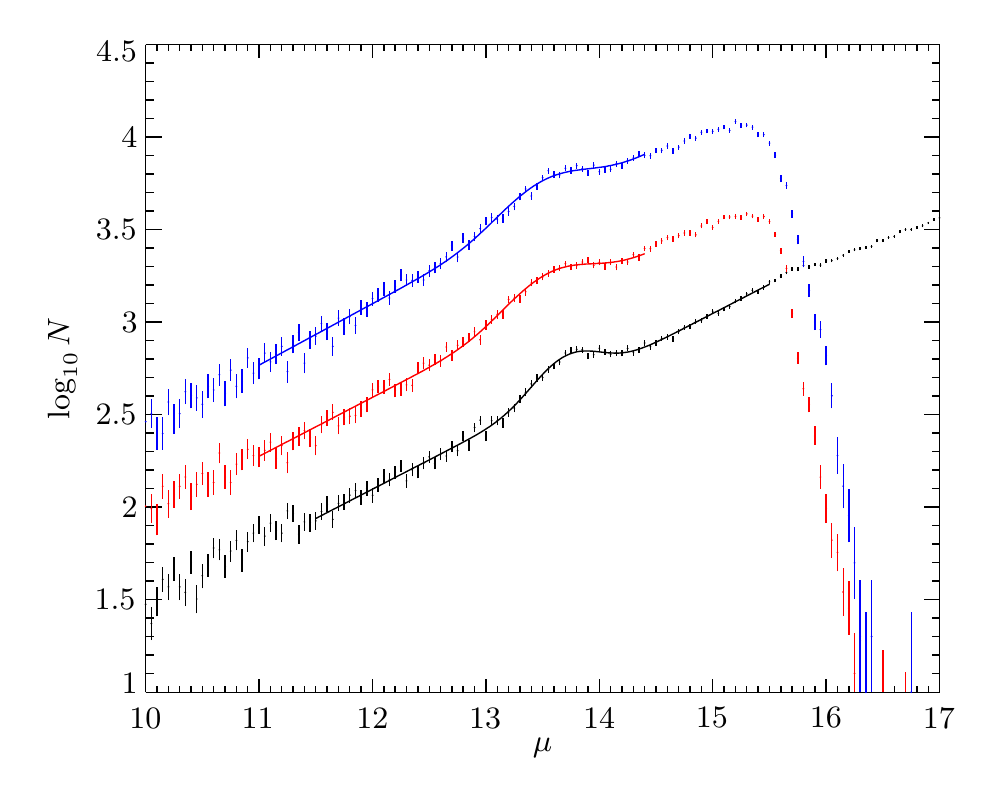}
\caption{Example histograms of distance modulus, $\mu$, calculated using equations \ref{eq:mujhk} and \ref{eq:muglimpse}, assuming that all stars are red clump stars. We show \Kband (black), \mone (red) and \mtwo (blue). We also show the fits to the histograms made using \autoref{eq:fiteq} as the solid lines. For legibility the \mone, and \mtwo bands are offset by $\log N=$ 0.5 and 1  with respect to the axis. This example field has its center at $l=18.5\dg$ $b=0.9\dg$ and has size $\Delta l=1\dg$, $\Delta b=0.3\dg$. \label{fig:examplefield}}
\end{figure}

\begin{figure*}
\centering
\includegraphics[width=0.75\linewidth]{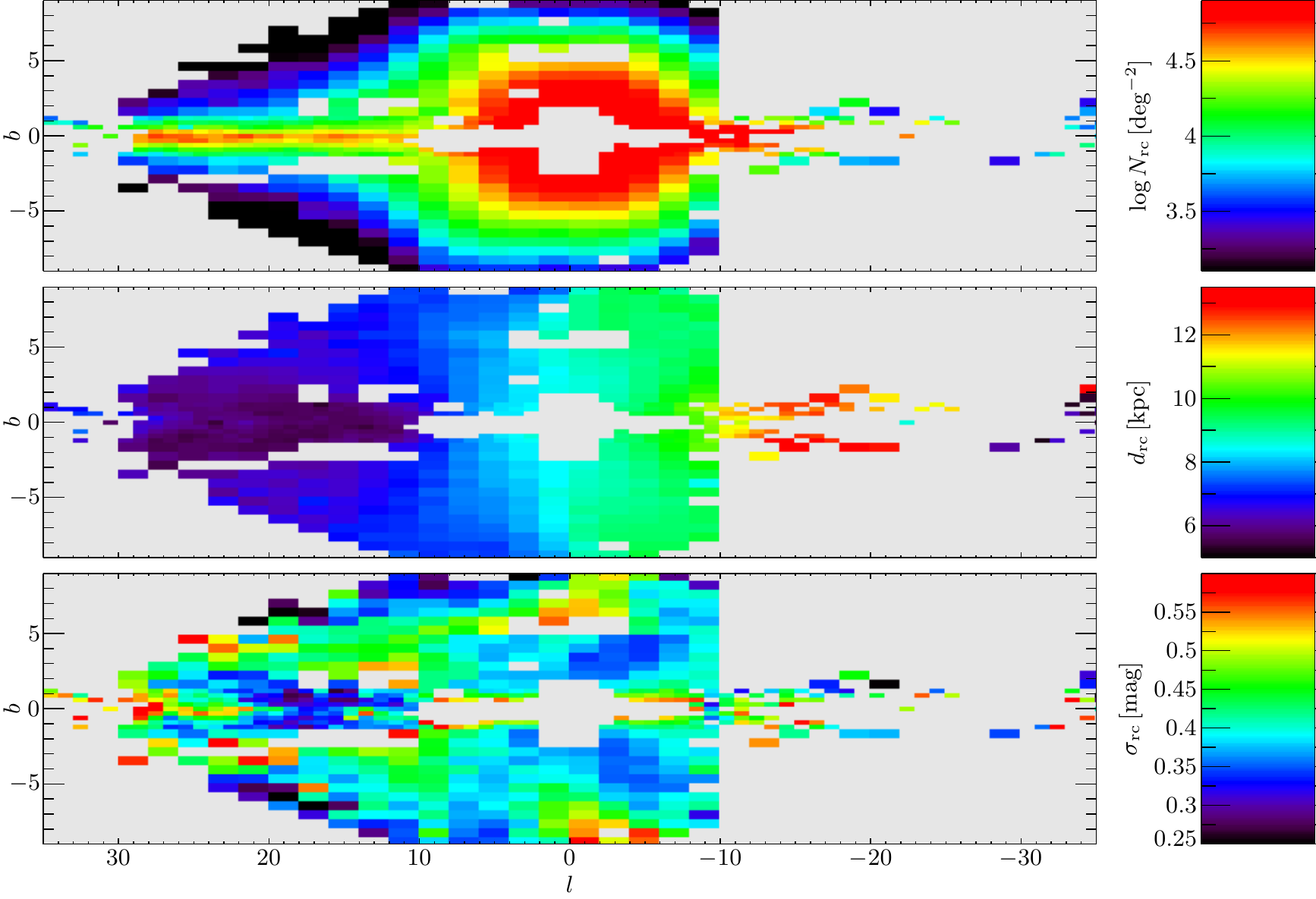}
\caption{Parameters of the Gaussian fits to red clump stars in the \Kband using \eqn{fiteq}. In the top panel we show the number of red clump stars $N_{\rm rc}$. In the middle panel we show the distance to the fitted peak of the red clump stars, $d_{\rm rc}$ (calculated from $\mu_{\rm rc}$). In the lower panel we show the dispersion in magnitudes of the fit to the red clump, $\sigma_{\rm rc}$. Grey regions are those where no well localised fit to the RCG distribution was possible, either because the surveys where not deep enough in that field, or the fitted dispersion was too large to represent RCGs towards the Galactic centre. \label{fig:kfits}}
\end{figure*}

\subsubsection{Catalogue Construction}

To utilise the star-by-star extinction correction methods we require band matched catalogs to calculate the reddening of each individual star. UKIDSS and 2MASS provide band matched catalogues while for the VVV survey we match between the $H$ and \Kband DR2 catalogs using a radius of 1\arcsec.

In combining the surveys we preferentially use VVV data when available, falling back to UKIDSS and finally 2MASS. The resultant footprint of the surveys is shown in the upper panel of \fig{extcormap}. In the lower panel of \fig{extcormap} we show the surface density of stars in the extinction corrected \Kband of our combined data. The long bar corresponds to the non-axisymetric in-plane enhancement of stars at positive longitudes.

\subsection{GLIMPSE data}

The longer wavelength GLIMPSE has lower dust extinction and we use it to cross check the \Kband data where possible. We use the $H-[4.5\mu]$ colour to correct for extinction corresponding the RJCE method \citep{Majewski:11}:
\begin{align}
\mu_{3.6\mu} =  [{3.6\mu}]&  - \frac{A_{3.6\mu}}{E(H-[4.5\mu])} \left[ (H-[4.5\mu] ) \right.  \nonumber \\ & \quad \left. - (H-[4.5\mu])_{\rm RC} \right] - M_{3.6\mu,{\rm RC}} \label{eq:muglimpse} \\
\mu_{4.5\mu} =  [{4.5\mu}] & - \frac{A_{4.5\mu}}{E(H-[4.5\mu])} \left[ (H-[4.5\mu] ) \right. \nonumber \\ & \quad\left. - (H-[4.5\mu])_{\rm RC} \right]- M_{4.5\mu,{\rm RC}} ~, \nonumber
\end{align}
where we adopt $\frac{A_{3.6\mu}}{E(H-[4.5\mu])}=0.31$ and $\frac{A_{4.5\mu}}{E(H-[4.5\mu])}=0.23$ corresponding to the measurements of \citet{Zasowski:09} in the range $15\dg <l<20\dg$ together with $A_H/A_\K=1.73$.
Measuring the colour of RCGs using the Padova isochrones as in \autoref{sec:kextcor} results in $[3.6]-\K=-0.05,\, [4.5]-\K=0.04$ and therefore $M_{3.6\mu,{\rm RC}}=-1.77$ and $M_{4.5\mu,{\rm RC}}=-1.68$.

To utilise the $H-[4.5\mu]$ colour for extinction correction we must band match the GLIMPSE data to surveys providing $H$-band measurements.  We have band matched to UKIDSS and VVV $H$-band data, using a radius of 1\arcsec, while GLIMPSE already provides matches to 2MASS. We utilise the deeper UKIDSS and VVV $H$-band data over 2MASS to correct for extinction in GLIMPSE when possible. This is particularly important close to the galactic plane where otherwise 2MASS would limit the depth of the GLIMPSE data to brighter than the clump.

In \autoref{fig:ext_ext} we compare the \Kband extinction calculated using the RJCE method (\autoref{eq:muglimpse}) to the value determined from the $H-\K$ reddening used by \autoref{eq:mujhk}. We find good agreement between the measurements using the two methods. The mean difference is less than $0.05\mags$ at all extinctions. The standard deviation in the difference between the estimates is typically less than $0.1\mags$ at magnitudes corresponding to long bar red clump stars. The majority of this dispersion is due to the photometric error in the \mtwo band which contributes $0.07\mags$ at low extinction. We have performed the same comparison as a function of galactic longitude and latitude and found no significant differences. This confirms that the extinction law does not vary significantly across the region considered, and that the different surveys are consistent.

Finally we construct magnitude distributions in fields covering $|l| <  40\dg$, $|b| <  9\dg$. We use two grids: close to the galactic plane at $|b| < 1.2\dg$ we use fields of size $\Delta l = 1\dg$, $\Delta b = 0.3\dg$. Away from the galactic plane at $|b| > 1.2\dg$, where the number count surface densities are lower, we use $\Delta l = 2\dg$, $\Delta b = 0.6\dg$.

In \autoref{fig:examplefield} we show, for an example field, histograms calculated using equations \ref{eq:mujhk} and \ref{eq:muglimpse}. In this field RCG stars in the bar are visible as the `bump' of stars at $\mu\approx 13.5$ above the smooth background of non-RCG bar stars.

\section{Gaussian Red Clump Fits}
\label{sec:rcfits}

\begin{figure*}
\includegraphics[width=0.95\linewidth]{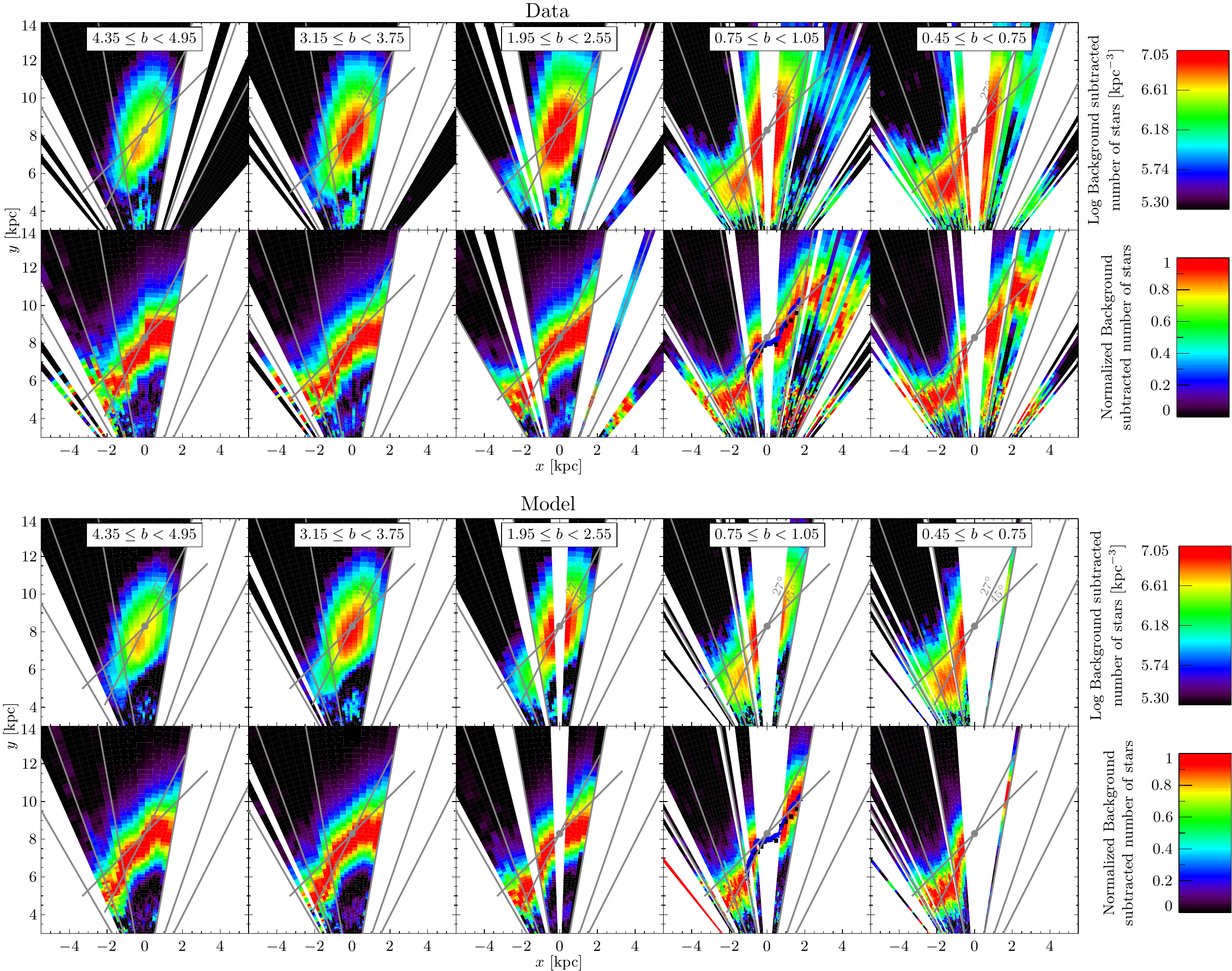}
\caption{\label{fig:kaboverho} In upper half of the figure we show selected slices of the Milky Way bar  viewed from above using the \Kband. In the lower half of the figure we show the two component model  constructed in \autoref{sec:fitrho} observed in the same manner. Data are the raw histograms constructed using \autoref{eq:mujhk} with the fitted exponential representing non-red clump stars subtracted. The data is divided by the volume of each bin to approximately convert the histograms to densities in ${\rm kpc}^{-3}$. In the top row of plots we show these densities on a log scale. The lower rows are the same data, but each line-of-sight is normalised to its peak. On the $0.75 \le b < 1.05$ slice we also plot the red clump distances found by \citet{Nishiyama:09} in blue, and \citet{gonzalez:11a} in black, using our adopted $m_{K,{\rm RC}}=-1.72$. To reduce noise the histograms are smoothed using boxcar smoothing with a width of 0.15\,mag, corresponding to three histogram bins. Red lines show longitudes of $l=-30,-20,-10,10,20,30\dg$ and lines at $\alpha=27\dg$ and $45\dg$ to the sun line-of-sight with half length 4.7\kpc.} 
\end{figure*}

To the magnitude distribution in each field we attempt to fit a Gaussian representing the red clump together with a exponential background \citep{Stanek:95,Nataf:13}:
\begin{equation}
N(\mu) = \frac{N_{\rm RC}}{\sigma_{\rm RC} \sqrt{2 \pi}} \exp \left[ -\frac{1}{2} \left( \frac{\mu - \murc}{\sigma_{\rm RC}}  \right)^2 \right] + A \exp B ( \mu - \murc )~. \label{eq:fiteq}
\end{equation}
We assume that the error is the Poisson error on the number in each magnitude bin and fit this equation using $\chi^2$. Each fit is performed using an Markov Chain Monte Carlo (MCMC) of 20,000 steps and flat priors, starting from the maximum likelihood position. 

In regions covered by UKDISS or 2MASS we fit over the range $11.5 < \mu_K < 15.5 $, in the VVV bulge region we use the range $12 < \mu_K < 16$, and in the VVV disk region we use the range $12 < \mu_K < 16.5$. These limits were determined by visual inspection. The GLIMPSE $\mu_{3.6\mu}$ and $\mu_{4.5\mu}$ data is shallower and in this case we extend the fit to 10.5 at the bright end, while the faint end is empirically determined in each field: In each field we estimate the distance modulus at which the counts drops to 50\% of the peak due to incompleteness and fit to a maximum $\mu$ of 0.75\mags brighter than this. Examples of these fits are shown in \fig{examplefield}.

\begin{figure}
\includegraphics[width=\linewidth]{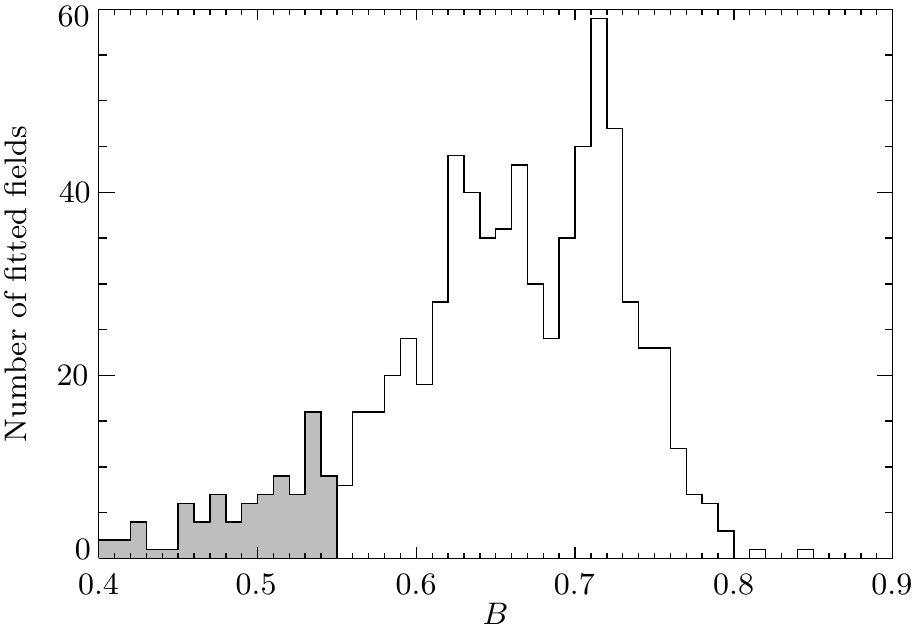}
\caption{\label{fig:Bhistogram} The fitted slope of the background of stars which are not RCGs defined though \autoref{eq:fiteq}. The grey region are fields which are subsequently rejected for having $B<0.55$ and therefore being significantly incomplete. The majority of these are crowded central bulge fields. If all stars are giants then because of the power law slope of the luminosity function on the giant branch we would expect this parameter to lie in the range $0.6 \protect\lesssim B \protect\lesssim 0.78$ \citep{Mendez:02}.} 
\end{figure}

The resultant fitted parameters, $N_{\rm RC}$, $\murc$, and $\sigma_{\rm RC}$, for the \Kband are shown in \fig{kfits}. In this plot we only consider the fit as good if the red clump is detected at $>3\sigma$ \ie $N_{\rm RC}$ is at least three times larger than its error, and the background slope $B > 0.55$. The second criterion is equivalent to rejecting fields which are significantly incomplete at the faint magnitude limit since in this case the slope of the fitted background is reduced. We also require that the best fitting red clump magnitude lies within the range of magnitudes fitted, and that the dispersion of the fitted red clump, $\sigma_{\rm RC}<0.6$ since visually fits that do not meet this criterion appear spurious. Finally we exclude all fields at $l<-10\dg$ without VVV data since 2MASS data is not deep enough the detect the clump in this region and the few fits which passed the previous criteria were visually rejected. We show in \autoref{fig:Bhistogram} the histogram of fitted values of $B$ for fields that pass the other selection criteria.

Several features of the fits in \fig{kfits} are worth noting:
\begin{inparaenum}
\item the red clump is detected and fitted up to $|b| \approx 5\dg$ at $l > 10\dg$ out to $l \approx 20\dg$, outside the traditionally defined bulge region. In this region the fitted distance at the same $l$ is close to the distance fitted in the Galactic plane, with an increased dispersion.
\item Fits no longer pass our criteria for a well fitted red clump at $l \gtrsim 30\dg$. This is similar in longitude to that identified as the end of the long bar in the literature \citep[\eg][]{LopezCorredoira:06,CabreraLavers:07}. We address this in greater detail in \autoref{sec:barlength} and \autoref{sec:barlengthdiscuss}.
\item The feature visible from $l=-20\dg$ to $l=-25\dg$ in the surface density of stars in \autoref{fig:extcormap} is not visible here. This is either because it is composed exclusively of young stars which have not yet evolved to become clump stars, or is too extended along the line-of-sight to give a well localised  red clump. 
\item Along the minor axis of the bulge at $|b|>5\dg$ the fitted dispersion is larger than off-axis. This is a result of using one Gaussian to fit the wider split red clump \citep{Nataf:10,McWilliam:10} in this region.
\item In the Galactic plane at $l > 10\dg$ the scatter in fitted distance and dispersion are considerably smaller than in \citet{CabreraLavers:08}.
\end{inparaenum}

In \fig{kaboverho} we show the raw histograms but with the exponential background  of non-red clump stars subtracted \ie we subtract the fitted $A \exp B ( \mu - \murc )$ from each histogram. This background subtracted histogram for each line-of-sight is then plotted as if viewed from the north Galactic pole. In addition we divide the number in each histogram bin by its volume, where the volume of each equally spaced bin in magnitude increases with distance like $\propto d^3$. This is equivalent to plotting the density assuming that red clump stars are perfect standard candles. From these plots it is already evident that the bar more closely matches the bar angle shown in the figure of $\alpha=27\dg$ \citep[\eg][]{Wegg:13} than $\approx 45\dg$ \citep[\eg][]{CabreraLavers:08}, although the data at low latitudes shows a possible curvature towards the end of the bar in a leading sense similar to that suggested by \citet{MartinezValpuesta:11}. 

We have performed the same analysis with the \mone and \mtwo bands. The equivalent plots to Figs. \ref{fig:kfits} and \ref{fig:kaboverho} are shown figures \ref{fig:glimpseextcormap}-\ref{fig:3_6aboverho} in \autoref{glimpseappend}. The GLIMPSE data for $b>0.45\dg$ are consistent with the \Kband data. In particular, they independently support a bar angle near $27\dg$. At $b<0.45\dg$ GLIMPSE does not provide sufficient completeness for RCGs for a good comparison. 

In summary we find that the long bar matches a bar angle of 27\dg more closely than 45\dg in both the \Kband and the GLIMPSE data, and that the long bar is detected to $|b|=5\dg$ outside the bulge at $l<20\dg$.

\section{Long Bar Vertical Profile: Longitude Slice Fits}
\label{sec:slicefits}

We estimate the scale height by performing fits to the number of red clump stars as a function of latitude. We treat each longitude slice independently and fit the $N_{\rm rc}$ found in from \eqn{fiteq} to an exponential:
\begin{equation}
N_{\rm RC}(b) = \frac{ \Sigma_{\rm RC} }{2 b_1} \exp \left( -\frac{\left|b-b_0\right|}{b_1} \right) ~, 
\label{eq:scaleheight1}
\end{equation}
where $b_0$ is an offset and $b_1$ the exponential scale height. In fields from $15\dg \lesssim l \lesssim 25\dg$ we find that a single exponential is a poor fit to our data, which extends to higher latitudes than previous investigations. We therefore additionally fit a double-exponential:
\begin{equation}
N_{\rm RC}(b) = \frac{\Sigma_{{\rm RC},A}}{2 b_{1,A}} \exp \left( -\frac{\left|b-b_0\right|}{b_{1,A}} \right) + \frac{\Sigma_{{\rm RC},B}}{2 b_{1,B}} \exp \left( - \frac{\left|b-b_0\right|}{b_{1,B}} \right)  ~.
\label{eq:scaleheight2}
\end{equation}

\begin{figure}
\includegraphics[width=\linewidth]{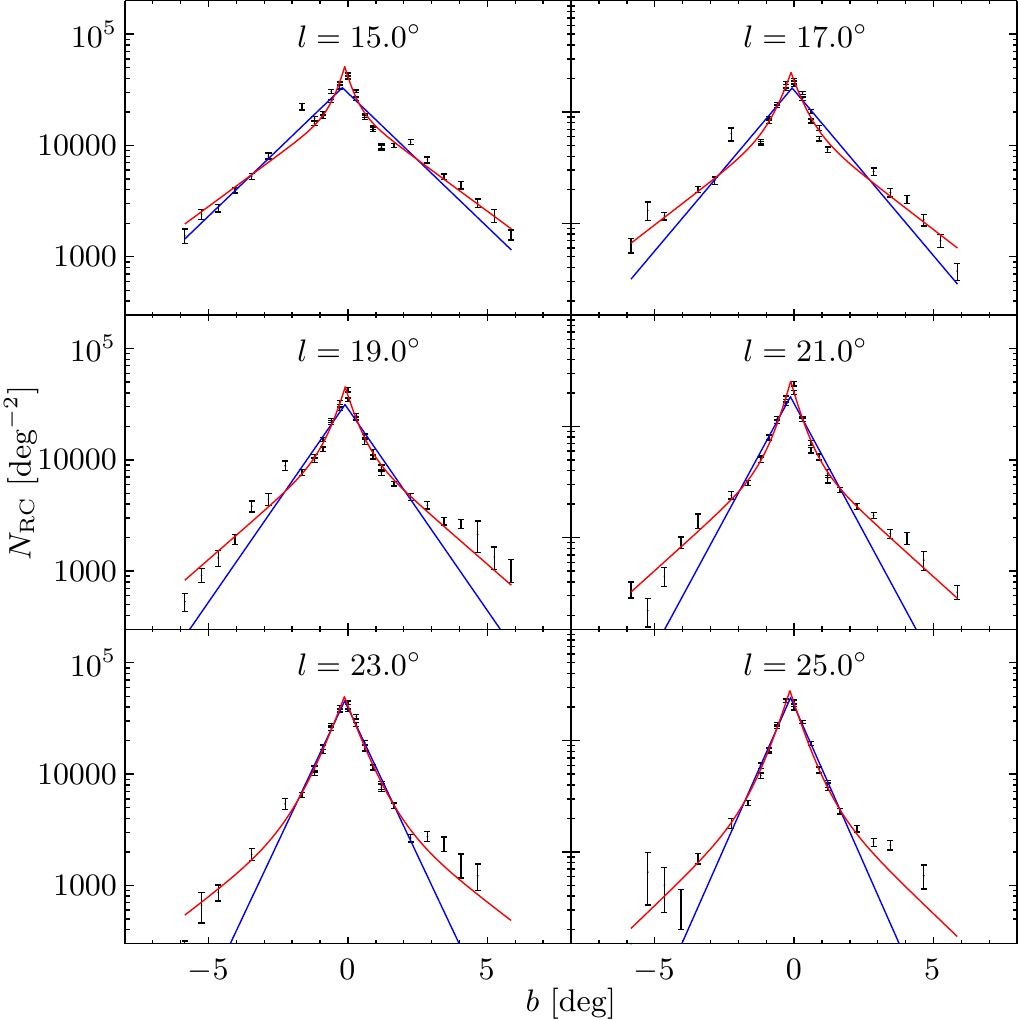}
\caption{The vertical profile of the surface density of red clump stars identified in the \Kband in several longitude slices. The blue curve is the best fitting single exponential to the \Kband, while the best fitting double exponential is the red curve. Error bars show the statistical error resulting from fitting \autoref{eq:fiteq} to each field. \label{fig:scaleheightexamples}} 
\end{figure}

\begin{figure}
\includegraphics[width=\linewidth]{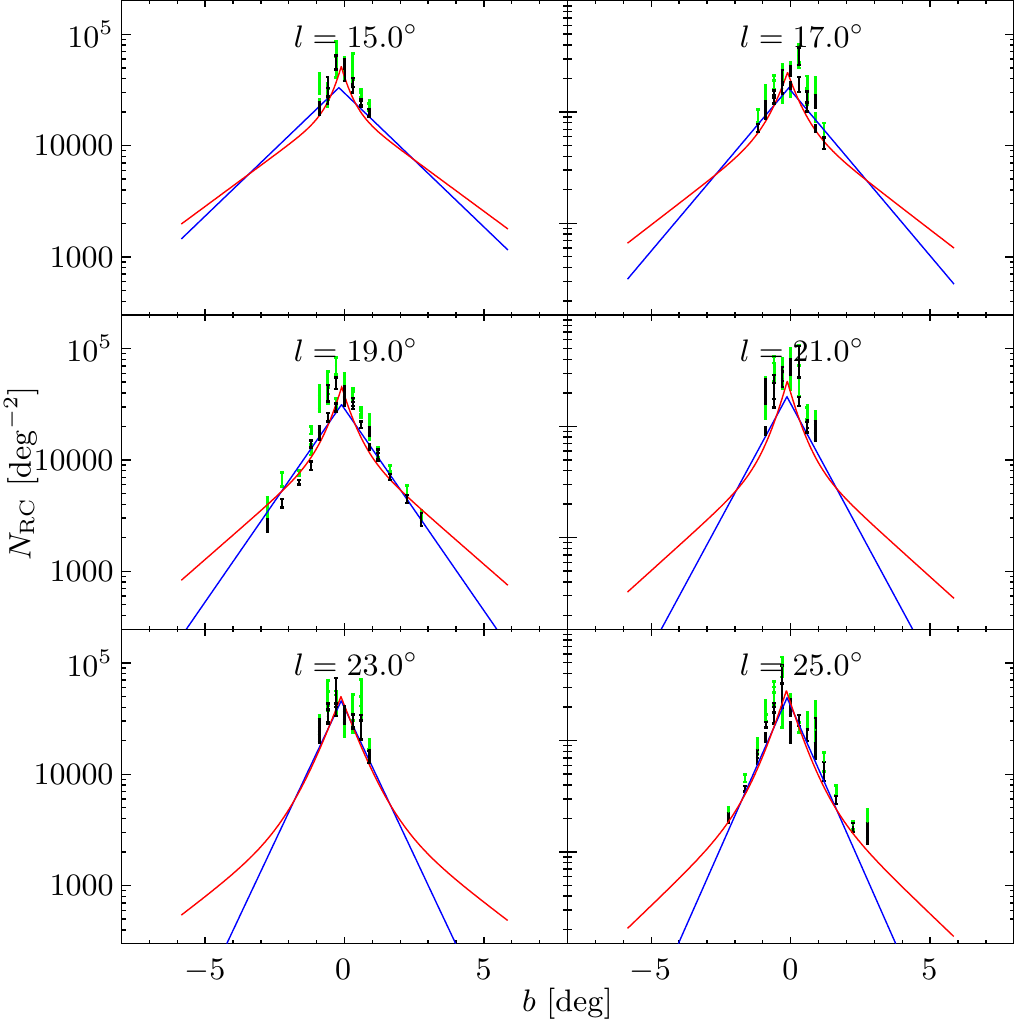}
\caption{The vertical profile of the surface density of red clump stars in several longitude slices in GLIMPSE data compared to the profiles fitted to the \Kband. In black we show the surface density of red clump stars identified in the \mone band and in green those identified in the \mtwo band. The blue curve is the best fitting single exponential to the \Kband, while the best fitting double exponential is the red curve. The curves have not been renormalised. Error bars show the statistical errors resulting from fitting \autoref{eq:fiteq} to each field. \label{fig:glimpsescaleheightexamples}} 
\end{figure}

Examples of these fits for the long bar region, where we find evidence for two scale heights, are shown in figures \ref{fig:scaleheightexamples} and \ref{fig:glimpsescaleheightexamples}. In \autoref{fig:scaleheightexamples} we show the \Kband data and the fits to these profiles. In \autoref{fig:glimpsescaleheightexamples} we show the GLIMPSE data but with the fits to the \Kband overplotted. The GLIMPSE data do not extend sufficiently far from the galactic plane to verify the double exponential fits, however the agreement in the region close to the galactic plane is an important check since this is the region where the extinction is greatest and the longer wavelength GLIMPSE data is affected less.

\begin{figure}
\includegraphics[width=\linewidth]{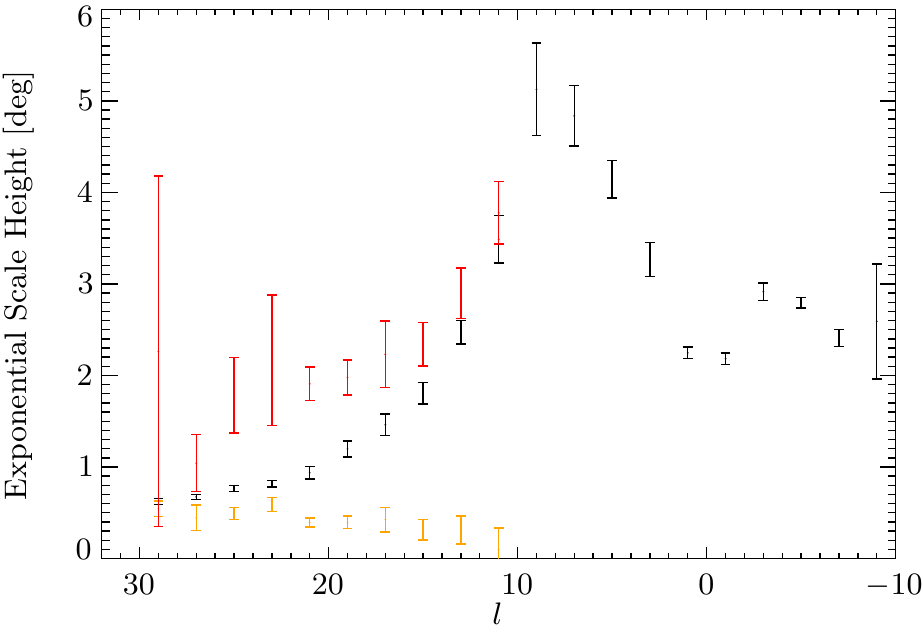}
\caption{Scale height in degrees in each longitude slice measured by fitting the single and double-exponential models to $N_{\rm rc}$ from \eqn{fiteq}. In black we show the results of fitting one exponential, while in red and orange we show the results of fitting a double exponential. \label{fig:expscaleheight}}
\end{figure}

The resultant scale heights of these fits are shown in \autoref{fig:expscaleheight}. The black points show the exponential scale height of RCGs. The scale height increases from the bulge minor axis to a peak near the end of the Box/Peanut region, before continuously decreasing through the long bar region to $<1\dg$ near the bar end.
In the long bar region where we find that a double-exponential fits the data better the scale heights are typically $\approx 0.5\dg$ and $\approx 2\dg$. At $l=20\dg$ the bar lies at a distance of $\approx 5.2\kpc$ and these therefore correspond to scale heights of $\approx 45\pc$ and $\approx 180\pc$. The $\approx 180\pc$ scale-height component is similar in scale height to the thin disk in the solar neighbourhood and we therefore analogously refer to it as the \emph{thin bar}. This is appropriate since the scale height of thin disks in edge-on external galaxies varies weakly with position \citep{vanderKruit:81,deGrijs:97}. We refer to the significantly thinner $\approx 50\pc$ component as the \emph{super-thin bar}, both due to its small scale height and in analogy to super-thin components found in edge-on external galaxies \citep{SchechtmanRook:13}.

\begin{figure}
\includegraphics[width=\linewidth]{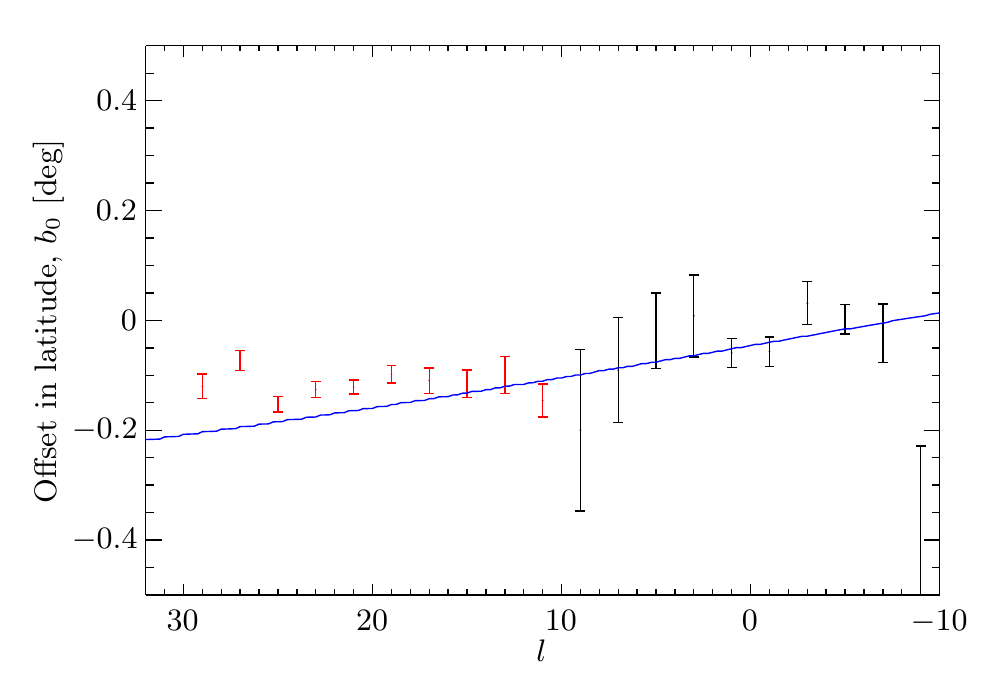}
\caption{Longitude offset in degrees in each longitude slice measured by fitting the single and double-exponential models to $N_{\rm rc}$ from \eqn{fiteq}. In black we show the results of fitting one exponential, while in red we show the results of fitting a double exponential. In blue we show the expected offset for a geometrically thin bar with bar angle 27\dg, the Sun lying 25\pc above the Galactic plane, and \sgra in the physical Galactic mid-plane at $b=-0.046\dg$ \citep{Goodman:14}.   \label{fig:offset}}
\end{figure}

We find that the fitted profiles are not symmetric about $b=0$ but instead require an offset. In \autoref{fig:offset} we show the fitted offset which is typically $b_0 \approx -0.1\dg$, corresponding to an offset of $\approx 14\pc$ at 6\kpc. For comparison, we can predict the latitude offset of the physical Galactic mid-plane as a function of distance and longitude, with two assumptions: that the Sun lies 25\pc above the galactic plane \citep{Maiz:01,Chen:01,Juric:08} and that \sgra lies in the Galactic plane with $b=-0.046\dg$, so that the physical Galactic mid-plane is slightly tilted with respect to the $b=0$ plane \citep[Fig. 2 of][]{Goodman:14}. We show in blue the resultant offset if the bar is geometrically thin along the line-of-sight with bar angle 27\dg. This simple prediction agrees well for $l\lesssim 15\dg$. It slightly overestimates the offset at $l\gtrsim 15\dg$ although typically by $<0.05\dg$. This could be explained by the apparent bar density peak lying behind the geometrically thin prediction for these longitudes, as is also seen in \autoref{fig:kaboverho}. We conclude that the long bar, and barred bulge is consistent with lying in the physical mid-plane of the galaxy to within approximately 0.05\dg in latitude, or around 5\pc for these distances. This could equally be seen as evidence supporting the assumptions that \sgra lies in the Galactic mid-plane, and the sun lies 25\pc above the mid-plane.   

\begin{figure}
\includegraphics[width=\linewidth]{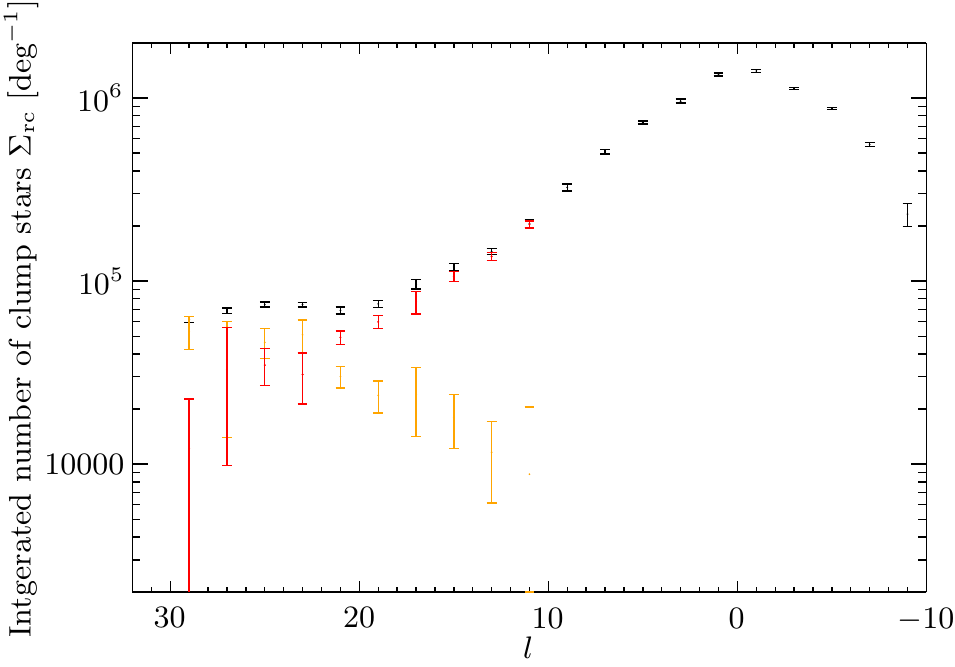}
\includegraphics[width=\linewidth]{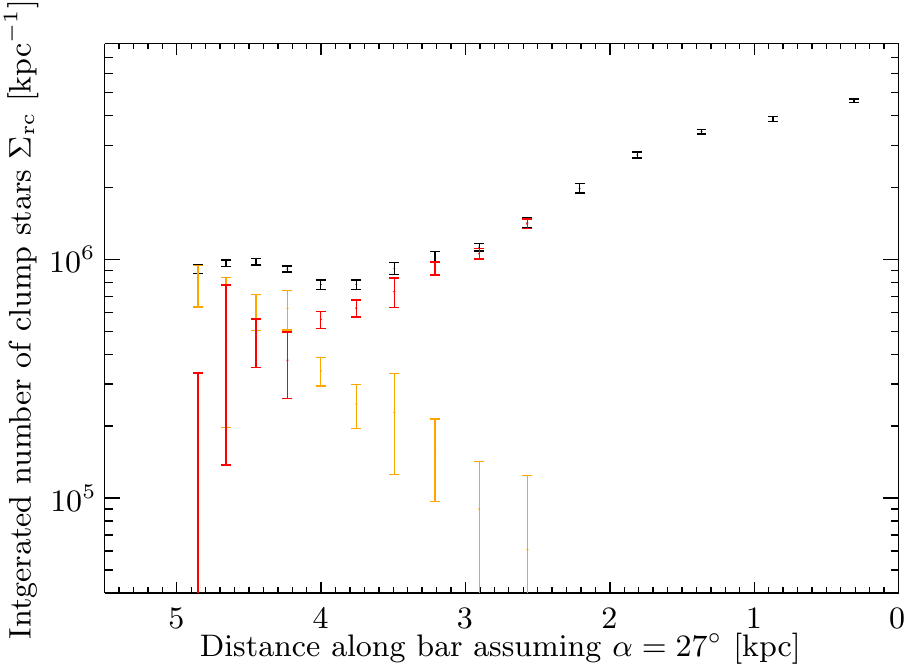}
\caption{In the top plot we show the integrated number of RCGs in the bar as a function of longitude, $\Sigma_{\rm rc}$, defined through equations (\ref{eq:scaleheight1}) and (\ref{eq:scaleheight2}). In black we show the result from fitting a single exponential (\autoref{eq:scaleheight1}). In red and orange we show the results from fitting a a double exponential (\autoref{eq:scaleheight2}) in the region where this is a significantly better fit. In the lower panel we show the same results with longitude converted to distance along the major axis of the bar assuming that it is geometrically thin along the line-of-sight and lies at bar angle $\alpha=27\dg$. 
\label{fig:sigmarc} }
\end{figure}

In \fig{sigmarc} we show the number of stars integrated over $b$ as a function of longitude and the integrated number of stars converted to a number of stars along the bar major axis, assuming a thin bar with angle $\alpha=27\dg$.
While the errors are large, the number of stars in the thin bar component in these plots decreases along the bar in a manor consistent with an exponential. Fitting the lower panel of \autoref{fig:sigmarc}, where we assume that the bar is thin and lies at $27\dg$ to the sun line-of-sight gives the exponential scale length of this component to be $1.5\kpc$, somewhat shorter than the thin disk scale length of $2-3\kpc$ (\eg  2.5\kpc: \citealt{Binney:97}, 2.4\kpc: \citealt{Bissantz:02}, 2.6\kpc: \citealt{Juric:08}, 2.15\kpc: \citealt{Bovy:13}). 
The number of stars in the super-thin component however increases outwards. We therefore propose that the short scale height component corresponds to stars formed more recently  predominantly towards the bar end, that have not experienced a large amount of vertical heating since birth. For reference, to have formed RCGs the stars must still be $\gtrsim 500\Myr$ old but galaxies with ongoing star formation have a strong bias to younger RCGs with a peak of age $\sim 1\Gyr$ \citep{Girardi:01,Salaris:02}. The larger scale height component would then correspond to old red clump stars analogous to old thin disk stars in the solar neighbourhood. We discuss this scenario further in \autoref{sec:discuss} after our more detailed modelling. 

We also note that both the scale height and the total number of red clump stars transitions smoothly from the bulge region ($|l|<10\dg$) to the long bar region ($10<l<30\dg$), providing further evidence that the two structures are not distinct.

\section{Best Fitting Parametric Density Models}
\label{sec:fitrho}

In this section we construct bar/bulge models that match the magnitude distributions and describe their key features such as the resultant bar angle, bar length and bar mass. Our approach is to model the stellar density, and convolve this with suitable luminosity functions to produce model magnitude distributions. The stellar density is then adjusted until the model magnitude distributions match the extinction corrected magnitude distributions. We again focus on the \Kband because of the wider and deeper coverage than GLIMPSE. We fit the magnitude distributions over the range $11 \le \mu_K \le 15$; this range corresponds to $1.6\kpc<r<10\kpc$ and therefore contains red clump stars in the bulge and long bar. Their nature as an approximate standard candle, together with their abundance, provides the statistical power in the method. 

We model only the density and do not attempt to build dynamical models. In the future we will use the observational constraints from this work together with the made-to-measure method to construct self-consistent models whereby the stellar density generates the potential in which the stars orbit, similar to \citet{Portail:15}. In principle fitting a non-parametric model similar to \citet{Wegg:13} would be preferable. Unfortunately the signal-to-noise of RCGs relative to the background of non-RCGs is significantly smaller in each field in the long bar than in the bulge region. The parametric models are fitted all fields simultaneously, thus improving the bar signal. The planned made-to-measure modelling would again be non-parametric, but would connect different fields by requiring that the model be dynamically self-consistent. We regard the parametric models constructed in this section as a useful step in uncovering the properties of the data even though they do not statistically match the data perfectly.
 
Our model density consists of three distinct parts: 
\begin{enumerate}[leftmargin=.15in]
\item An N-body model taken from \citet{Portail:15}. These N-body models were adjusted to fit the density and kinematics in the galactic bulge at $|l| < 10\dg$ and therefore fit this region well. As in \citet{Portail:15} the model is placed so that the bar is at an angle of $\alpha=27\dg$.

We use an SPH cubic spline kernel to evaluate a smooth density field from the N-body model. For clarity we repeat the formulae here following the notation of \citet{Hunt:12}. We use the cubic spline kernel
\begin{equation}
	W(s,h)=\frac{8}{\pi h^3}\times
	\begin{cases}
		1-6(s/6)^2+6(s/h)^3 & \mbox{if } 0 \leq s/h < 1/2\, , \\
		2(1-s/h)^3 & \mbox{if } 1/2 \leq s/h  < 1 \, , \\
		0 & \mbox{otherwise.} 
	\end{cases}
\end{equation}
The density at a point $\bm{x}_i$ is then given by
\begin{equation}
	\rho(\bm{x}_i)=\sum_{j=1}^N m_j W( | \bm{x}_i - \bm{x}_j|, h_j)
	\label{eq:sphrho}
\end{equation} 
where the sum runs over all N-body particles, $m_j$ is the mass of particle $j$. The smoothing length of particle $j$ is evaluated from the local density through
\begin{equation}
	h_j=\eta \left( \frac{m_j}{\rho(\bm{x}_j)} \right)^{1/3} ~,
	\label{eq:smoothlen}
\end{equation}
where $\eta$ is a parameter for which we have chosen $\eta=3$. Equations \ref{eq:sphrho} and \ref{eq:smoothlen} are solved iteratively until the difference between iterations is less than $10^{-3}$. The choice of both these parameters mirrors \citet{Hunt:12}.

\item We add parametric functions to represent the long bar component, the component of primary interest in this work. We choose the form \citep[similar to][with $c_\parallel=1$]{Freudenreich:98,Picaud:04,Robin:12}
\begin{align}
\rho = \frac{M_{\rm bar}}{4\pi x_0 y_0 z_0} \exp &\left(-\left[\left(\frac{x}{x_0}\right)^{c_\perp}+\left(\frac{y}{y_0}\right)^{c_\perp}\right]^{1/c_\perp} \right) \exp \left[ -\frac{z}{z_0} \right] \nonumber \\ &\times 
{\rm Cut}\left[ \frac{R-R_{\rm out}}{\sigma_{\rm out}} \right] {\rm Cut}\left[ \frac{R_{\rm in}-R}{\sigma_{\rm in}} \right]
\label{eq:barfunc}
\end{align}
where $x,y,z$ are right-handed galactocentric coordinates so that $x$ is orientated along the bar major-axis, $y$ along the bar intermediate-axis, and $z$ towards the north Galactic pole. $R=\sqrt{x^2+y^2}$ is the galactocentric radius. We use a Gaussian cutoff for the inner and outer edges of the bar:
\begin{equation}
{\rm Cut}(x)=
\begin{cases}
\exp(-x^2) & \mbox{if } x > 1 \\
1 & \mbox{if } x \le 1
\end{cases} ~ .
\end{equation}
Note that because of the Gaussian cutoffs the bar mass is less than the parameter $M_{\rm bar}$. The inner cutoff is required because the N-body model already fits the magnitude distributions in the central region well as it was tailored to fit the data there using the made-to-measure method in \citet{Portail:15}.

\item The Galactic disk scale length of the N-body model is $1.1-1.2\kpc$, significantly shorter than that of the Milky Way, $\approx 2-3\kpc$ (see \autoref{sec:slicefits}). This is true in many N-body bar models formed from initially unstable disks where the bar length is typically several disk scale lengths. That the long bar of the N-body model is less prominent than in the data is also related to the discrepancy in disk scale length, since moving outwards along the bar there are fewer stars than in the Milky Way. Here we do not attempt to model the disk to reconcile the discrepancy between the N-body disk and the Milky Way. Fitting the disk would be an involved task worthy of a separate work, while here we concentrate on features of the non-axisymmetric part of the density \ie the bar. Instead we add a best fitting additional exponential component to each modelled field magnitude distribution independently:
\begin{equation}
N_{\rm disk}(\mu_K) = N_{0,{\rm disk}} \exp \left[ b \left(\mu_K-13.5 \right) \right] ~.
\label{eq:disk}
\end{equation}
The result is that we are fitting the difference between the magnitude distribution and an exponential. We have verified that for the fields considered in this work, over the magnitude range considered, that the Besan\c{c}on Model \citep{Robin:03} predicts a nearly exponential magnitude distribution. Note that if the disk had a central hole that was significant for this work then the lack of RCGs in this region would result in a dip in the magnitude distributions in a similar manner to the peak caused by the bar. Since this is not seen in the fields fitted here we consider \autoref{eq:disk} adequate.

\end{enumerate}

In order to match the magnitude distributions in each field we convolve the resultant density from (i) and (ii) with a luminosity function to predict the number of stars as a function of $\mu_K$ before adding the disk from (iii) seperately. For clarity of notation we define the colour $\HK \equiv H-\K$, the corresponding reddening free colour $\MHK \equiv (H-\K)-E(H-\K)$, the reddening $\EHK \equiv E(H-\K)$, and $\extlaw \equiv A_\K/E(H-\K)$ which is a constant given by the extinction law.

Consider $\Phi(M_K,M_C)\,dM_K\,dM_C$ as the joint number of stars produced per unit mass between $M_K$ to $M_K+dM_K$ and $\MHK$ to $\MHK+d\MHK$ where capital $M$ denotes the absolute magnitude in the respective band. This two-dimensional colour-magnitude analogue to the luminosity function can be readily calculated by populating isochrones in a similar manner to the luminosity function. Define $N(\K,\HK)\,d\K\,d\HK$ as the number of stars observed in a pencil beam line of sight with solid angle $\omega$ jointly in the range $\K$ to $\K+d\K$ and $\HK$ to $\HK+d\HK$. Then the colour-magnitude version of the equation of stellar statistics is
\begin{align}
	N(\K,\HK)=\omega \int dr \, r^2 \rho(r) \Phi \big[ &\K - 5\log\left(r/10\pc\right) - A_\K(r) ~,  \nonumber \\ 
	& \qquad  \HK - \EHK(r) \big] \,  
\end{align}
where $A_\K(r)$ and $\EHK(r)$ are the \Kband extinction and the reddening in $C \equiv H-\K$ along the line-of-sight respectively. This can also be written in terms of distance modulus 
\begin{align}
	N(\K,\HK)= \int d\mu \, \Delta(\mu) \Phi \big[ &\K - \mu - A_\K(\mu) ~,~  \HK - \EHK(\mu) \big] \,   
\end{align}
where $\Delta$ is given by $\Delta(\mu) \equiv ({\rm ln} 10/5) \omega \rho r^3$ expressed in terms of $\mu$.
Changing variables from $(\K,\HK)$ to $(\mu_K,\MHK)$ allows us to calculate $N(\mu_K)\,d\mu_K$, the number of stars from $\mu_K$ to $\mu_K+d\mu_K$. This is the quantity which we compare to the data. Using \autoref{eq:mujhk} then
\begin{align}
N(\mu_K) &= \int  d\MHK \, N\Big[\mu_K + \extlaw \left( \HK - \MHK_{\rm , RC} \right) +M_{\K,{\rm RC}} ~,~ \MHK+\EHK(\mu)\Big]
 \nonumber \\ 
&=  \int d\mu \int d\MHK \, \Delta(\mu) \Phi \big[ \mu_K - \mu + \extlaw \left( \HK - \MHK_{\rm , RC} \right) \nonumber \\ 
&\qquad\qquad\qquad\qquad\qquad  +M_{\K,{\rm RC}} - A_\K(\mu) ~,~  \MHK \big] ~.
\end{align}
For an extinction law characterised by the constant $\extlaw\equiv \frac{A_\K}{E(H-\K)}$ the extinction can be calculated from the reddening,
\begin{equation}
A_\K(r) = \extlaw E_C = \extlaw \left( \HK - \MHK \right) ~.
\end{equation}
Therefore 
\begin{align}
N(\mu_K) &= \int d\mu \int d\MHK \,  \Delta(\mu) \Phi \big[ \mu_K - \mu +  \extlaw \left( \MHK - \MHK_{\rm , RC} \right) \nonumber \\ 
&\qquad\qquad\qquad\qquad\qquad + M_{\K,{\rm RC}} ~,~  \MHK \big]\\
& = \int d\mu \, \Delta(\mu) \Phi_{\mu_K} ( \mu_K - \mu ) \label{eq:conv} ~,
\end{align}
where $\Phi_{\mu_K}(M_{\mu_K})$ is the luminosity function in $\mu_K$ calculated from $\Phi(M_K,\MHK)$ through
\begin{equation}
\Phi_{\mu_K}(M_{\mu_K}) \equiv \int d\MHK \,	\Phi \big[ M_{\mu_K} +\extlaw \left( \MHK - \MHK_{\rm , RC} \right) +M_{\K,{\rm RC}} ~,~ \MHK \big] ~.
\end{equation}
This differs slightly from the luminosity function in $M_K$ because not all stars share the color of the red clump.  Because we use the extinction free magnitudes (equation \ref{eq:mujhk}) extinction does not enter in the final equation provided the survey is complete and the extinction law is correct. This is true of all stars, not just RCGs. Practically the convolutions in \autoref{eq:conv} are performed as parallel FFTs for speed. The exponential in $N(\mu_K)$ used to represent the disk described in (iii) is then added to the result of \autoref{eq:conv}.

The luminosity function $\Phi_{\mu_K}$ is calculated from the {\sc BASTI} isochrones \citep{Pietrinferni:04}. We fiducially use the $10\Gyr$,  $\alpha$-enhanced isochrones together with a \citet{Kroupa:01} IMF and the metallicity distribution measured by \citet{Zoccali:08} in Baade's window to generate the luminosity function. To  allow for the uncertainty in red clump magnitude, when fitted we allow a global shift in the luminosity function $\Delta K$.

We then fit by minimising the $\chi^2$ between the model and observed magnitude distributions assuming that the error is the Poisson error in observed number. If we label magnitude bins $i$ and fields $j$ then we minimise:
\begin{equation}
\chi^2 = \sum_{{\rm fields\,}i} ~ \sum_{{\rm magnitude\,bins\,}j} \frac{ \left( N_{{\rm model}\,i,j} - N_{i,j} \right)^2}{N_{i,j}} ~,
\end{equation}
where $N_{{\rm model}\,i,j}$ is the prediction of the model and $N_{i,j}$ the observed number. We use bins in magnitude of 0.05 which are sufficiently narrow to not artificially broaden the luminosity function. We fit only fields where the exponential slope fitted to the background in \autoref{sec:rcfits} was greater than $0.55$ since, as discussed in that section, smaller values indicate significant incompleteness at the faint end resulting in a smaller slope.

Throughout we place the Sun a distance $R_0=8.3\kpc$ from the Galactic center \citep{Reid:14,Sotiris:15} and $25\pc$ above the Galactic plane \citep{Maiz:01,Chen:01,Juric:08}, and \sgra in the physical Galactic mid-plane at $b=-0.046\dg$. This results in a slight tilt of the $b=0$ plane with respect to the physical Galactic mid-plane \citep[Fig. 2 of][]{Goodman:14}, but well within the errors when the Galactic coordinate system was originally defined. To move from axes $x,y,z$ aligned with the principle axes of the bar to Galactic coordinates we first rotate by bar angle $\alpha$ about the $z$-axis, then move the sun to $R_0=8.3\kpc$ and $z_0=25\pc$ from the Galactic centre, and finally tilt the Galactic coordinate system by 0.12\dg towards \sgra \citep{Goodman:14}. 

We describe the fitting as a three stage process: We first fit the N-body model to the central fields in \autoref{sec:bulgefit}, we then add an additional component to fit the insufficient long bar component in the N-body model in \autoref{sec:barfit}, and finally we add an additional component required to adequately fit the super-thin component towards the bar end in \autoref{sec:thinfit}. The model we regard as our best model is that with two parametric long bar components. We give the fitted parameters in \autoref{tab:parms} and the resultant physical quantities such as the bar length and the mass of the barred components in \autoref{tab:physicalparms}. 

\begin{figure*}
\includegraphics[width=\linewidth]{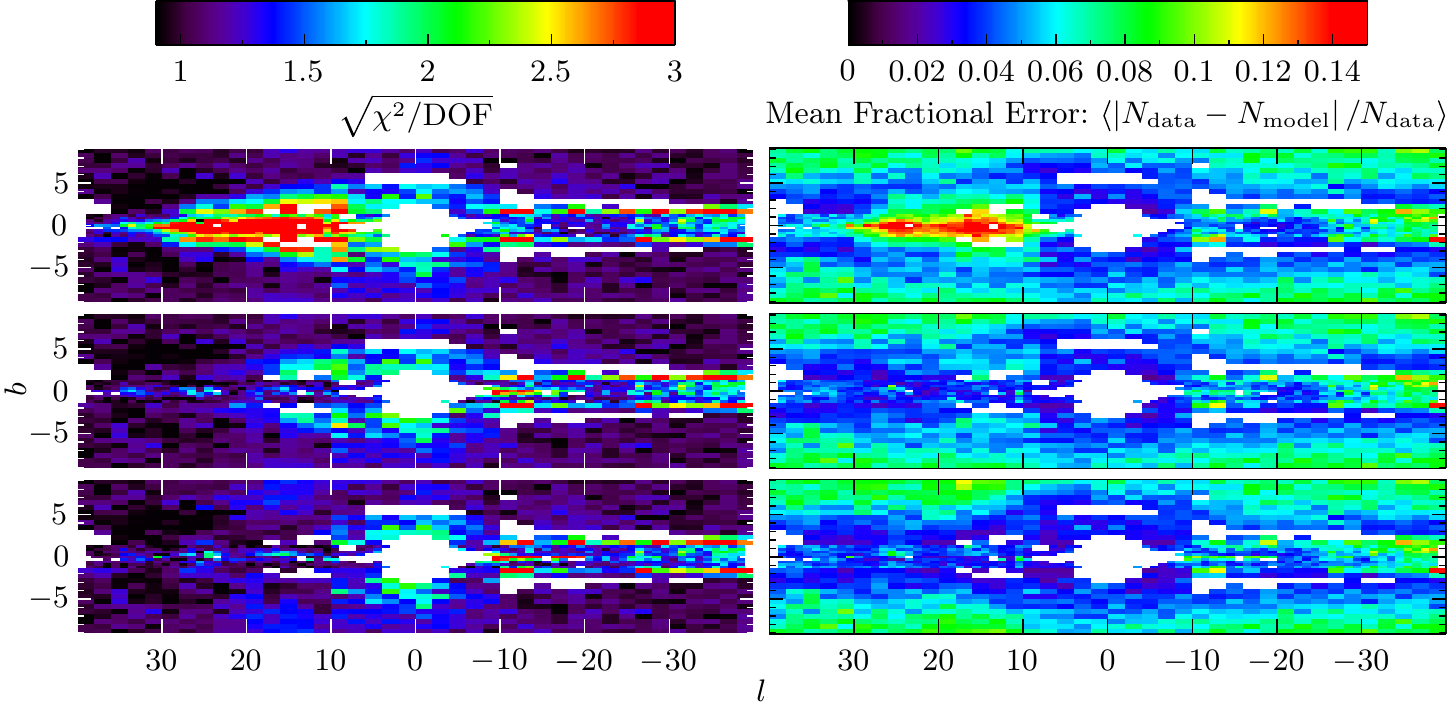}
\caption{In the left-hand three panels we show the reduced $\chi^2$ for the magnitude distribution in each field. In the right-hand three panels we show the mean absolute fraction error in each field over the range $13 \le K \le 15$. In the upper panel we show the fit using just the N-body model. In the middle panel we add an additional parametric bar component. In the lower panel add a second parametric component to fit the super-thin component at the bar end. White regions are those which were not fitted because they did not pass our completeness test.\label{fig:chi2_frac_all} }
\end{figure*}

\begin{figure}
\includegraphics[width=\linewidth]{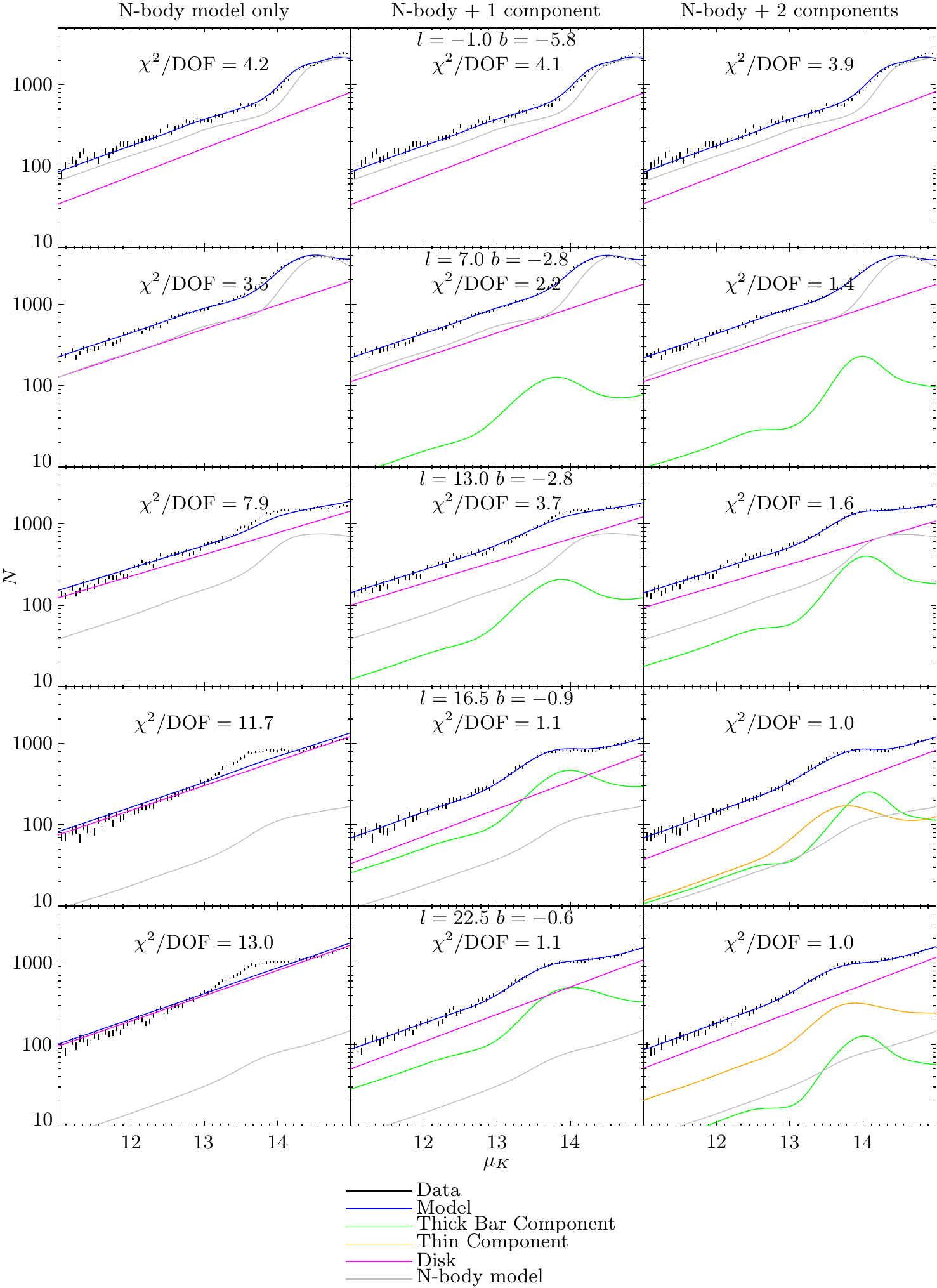}
\caption{Example fits to the magnitude distribution in selected fields. In the left column we fit only the N-body model  (and the additional field-by-field disk component \autoref{sec:bulgefit}). In the middle column we fit the N-body model together with a 1 component bar (\autoref{sec:barfit}).  In the right column we fit the N-body model together with a 2 component bar (\autoref{sec:thinfit}). The data with its associated Poisson error is plotted in black. The best fitting model is shown in blue. The model consists of a thin bar component (green), a super-thin bar component (orange), an exponential to represent the foreground disk (magenta) and the N-body model (grey).\label{fig:examplefieldfits}}
\end{figure}

\begin{table*}
\begin{minipage}{\linewidth}
\caption{\label{tab:parms} Parameters found when fitting model densities to the magnitude distribution. The parameters and the method of fitting is described in \autoref{sec:fitrho}. Note that $M_{\rm bar}$ differs from the physical bar mass give in \autoref{tab:physicalparms} because of the Gaussian cutoffs.}
\begin{tabular}{@{}l@{\hskip 0.05in} c@{\hskip 0.07in} c@{\hskip 0.07in}c @{\hskip 0.05in}c@{\hskip 0in} c@{\hskip 0.06in}c@{\hskip 0.06in}c@{\hskip 0.06in}c@{\hskip 0.06in}c@{\hskip 0.06in}c@{\hskip 0.06in}c@{\hskip 0.06in}c@{\hskip 0.06in}c@{\hskip 0.06in}c @{\hskip 0.05in}c@{\hskip 0in}}
\hline
& $\chi^2/{\rm DOF}$ & $\Delta K$ & $G$ & & $\log M_{\rm bar}/M_\odot$ & $\alpha\,(\deg)$ & $x_0\,(\kpc)$ & $y_0\,(\kpc)$ & $z_0\,(\kpc)$ & $R_{\rm out}\,(\kpc)$ & $\sigma_{\rm out}\,(\kpc)$ & $R_{\rm in}\,(\kpc)$ & $\sigma_{\rm in}\,(\kpc)$ & $c_\perp$ & \\
\hline
N-body model only & 2.06 & 0.17 & 1.47 &  &  - &  - &  - &  - &  - &  - &  - &  - &  - &  - & \\
N-body model + 1 component & 1.58 & 0.17 & 1.48 &  & 10.15 & 28.4 & 3.05 & 0.68 & 0.09 & 3.85 & 0.72 & 3.45 & 0.71 & 2.27 & \\
\multirow{2}{*}{N-body model + 2 components} &\multirow{2}{*}{1.54} &  \multirow{2}{*}{0.16} &  \multirow{2}{*}{1.45} &   \multirow{2}{*}{$\bigg\{$} & 10.87 & 29.1 & 0.82 & 0.23 & 0.20 & 4.37 & 0.11 & 3.26 & 0.56 & 3.98& \multirow{2}{*}{$\bigg\}$} \\
 & & & & & 9.97 & 30.0 & -32.66 & 1.88 & 0.04 & 3.36 & 0.76 & 7.09 & 2.24 & 1.67 &  \\
\hline
\end{tabular}

\end{minipage}
\end{table*}

\quad

\subsection{Bulge Fitting}
\label{sec:bulgefit}

To demonstrate and check the method we first fit the central bulge region, $\left| l \right| < 10\dg$. We take the model named M85 from \citet{Portail:15} as a fiducial fit. We fit for two global parameters: the normalisation of the model density $G$, and an offset in red clump magnitude $\Delta K$ from that assumed by the luminosity function constructed from the isochrones. 

We show in the top left panel of \autoref{fig:chi2_frac_all} the resultant $\chi^2$ value in each field across the entire range of longitudes, although only the central region was fitted. We also show in the top right panel of \autoref{fig:chi2_frac_all} the mean absolute fractional error in each field over a magnitude range covering the red clump.   Comparing these panels it can be seen although that the reduced $\chi^2$ is formally poor in the central fields the fractional error is extremely small. The larger $\chi^2$ here is due to the large numbers of stars and resultant small Poisson errors. The reverse is true at $|b|\gtrsim 8\dg$: the small number of stars results in larger statistical errors and therefore larger mean absolute fractional deviations, however the $\chi^2$ values demonstrate the fit is statistically good. It is evident that the model fits poorly in the fields close to the plane at $l>10\dg$. This is a result of the insufficient long bar component in the N-body model in comparison to the Milky Way. 

In the left hand column of \autoref{fig:examplefieldfits} we show some example fields and the resultant fits. Again it is clear that the fit in the bulge region is excellent, however the long bar is insufficient outside this central region. The fact that the N-body model together with the disk model fits well in the bulge region shows that the disk model is reasonable, since the N-body model was constructed to fit bulge-only deconvolved data.

We give our fitted parameters in \autoref{tab:parms}. The best fitted shift in red clump magnitude, $\Delta K$, is 0.17, suggesting that either $R_0$ is slightly less than our adopted $8.3\kpc$, or bulge red clump stars are slightly brighter than assumed in the luminosity function. Since our assumed bulge luminosity function places the red clump at $-1.59$, this would correspond to a fitted red clump absolute magnitude of $M_{K,{\rm RC}}=-1.75-5\log[R_0/8.3\kpc]$. This is brighter than the $M_{K,{\rm RC}}=-1.61$ measured from nearby Hipparcos stars \citep{Laney:12}. The normalisation of the model density, $G$, of $1.47$ depends on the N-body model chosen which differ in their dark matter content in the Bulge, and is also degenerate with the IMF. This is discussed in detail in \citet{Portail:15}.

\subsection{One Component Long Bar Fitting}
\label{sec:barfit}

\begin{table*}
\begin{minipage}{\linewidth}
\begin{center}
\caption{\label{tab:physicalparms} The derived physical parameters, bar mass and bar length, of the models with parameters fitted in \autoref{tab:parms} and described in \autoref{sec:fitrho}. }
\label{symbols}
\begin{tabular}{l c cccc c cccc}
\hline
& \multirow{2}{*}{$\bigg|$} & \multicolumn{4}{|c|}{Bar Mass $(10^9 M_\odot)$} & \multirow{2}{*}{$\bigg|$} & \multicolumn{4}{|c|}{Bar Half Length $({\rm kpc})$} \\
& & N-body & Thin Bar & Super-thin Bar & Total & & $L_{\rm drop}$ & $L_{m=2}$ & $L_{\rm prof}$ & $L_{\rm mod}$\\
\hline
N-body model only & & 11.0 & - & - &  11.0 & & 4.19 & 4.03 & 4.21 & - \\
N-body model + 1 component & & 11.1 & 8.8 &  - & 19.9 & & 5.50 & 5.44 & 5.65 & 4.87  \\
N-body model + 2 components & & 10.9 & 4.0 & 3.3 & 18.1 &  & 4.98 & 4.96 & 5.23 & 4.73 \\
\hline 
\end{tabular}

\end{center}
\medskip
The bar mass of the N-body model is 	 calculated by integrating the face-on surface density over all radii with the minor axis profile in the surface density subtracted. The thin and super-thin bar masses are the total mass in each component. The bar half lengths are as defined in \citet{Athanassoula:02} and all are measurement on the face on density map: $L_{\rm drop}$ is the radius at which the ellipticity drops fastest, $L_{m=2}$ is the radius at which the $m=2$ component of the face on image drops below 20\% of its maximum, and $L_{\rm prof}$ is the radius beyond which the major and minor axis density profiles agree within 30\%. Our threshold for $L_{\rm prof}$ is larger than the 5\% used by \citet{Athanassoula:02} which we found gave spuriously large bar lengths for the densities considered here. $L_{\rm mod}$ is the point at which the difference between the major and minor axis face-on surface densities falls to below $1/e$ of a fitted exponential long bar surface density profile. It has the advantage that it is independent of the axisymmetric disk. This measurement is spurious for the pure N-body model since it does not occur within the solar radius.
\end{minipage}
\end{table*}

We now add an additional parametrised function to represent the deficient long bar in the N-body model clear from \autoref{sec:bulgefit}. The functional form given in \autoref{eq:barfunc} is equivalent to the exponential shape function fitted in \citet{Robin:12}. We choose this form based on the results of \autoref{sec:rcfits} and \autoref{sec:slicefits} that the vertical structure is approximately exponential.  The resultant horizontal density profiles are elliptical in the case of $c_\perp=2$ and boxy for $c_\perp>2$. We have added an additional Gaussian inner cutoff function since the N-body model already fits well in the central region.

We fit this functional form, together with the N-body model and minimise $\chi^2$ in fields with $l>-10\dg$. The best fitting parameters are given as the second row of \autoref{tab:parms}. The reduced $\chi^2$ and mean absolution deviation are plotted in the second row of \autoref{fig:chi2_frac_all} from which it is clear that the resultant fit is greatly improved close to the plane in the region $10\dg<l<30\dg$. Note that the region near $5\dg<l<15\dg$ and $|b|\approx5\dg$ is not yet well fitted. The fitted bar angle is $\alpha=28.4\dg$ in agreement with the bar angle in the bulge region found by \citet{Wegg:13} of $\alpha=(27\pm2)\dg$. 

The additional mass associated with this component is $8.8\times10^{9}\msun$ for our luminosity function calculated from the BASTI isochrones for a 10\Gyr population with a \citet{Kroupa:01} IMF . If we also include the non-axisymmetric mass from the N-body model, calculated by integrating the face-on surface density over all radii with the face-on minor axis profile subtracted, then the total bar mass is $1.99\times 10^{10}\msun$. Note though that the mass derived here, which is constrained by the number of red clump stars, is degenerate with the IMF and population age. This is because the mass per red clump star varies in a similar manner to the mass-to-light ratio. Refitting changing from our fiducial Kroupa IMF to a Salpeter IMF \citep{salpeter:55} increase all masses by a factor of 1.43, while keeping all other parameters the same. Changing from a $10\Gyr$ to a constant star formation rate would reduce the mass by a factor of $2.0$. In this case since the luminosity function changes slightly the best fitting parameters change marginally from our fiducial values, however the physical properties of bar angle and length do not change significantly.Uncertainty in IMF and population age dominate the bar mass measured in this work and are larger than the differences between models or bar parameterisations.

\subsection{Two Component Fitting}
\label{sec:thinfit}

The adopted parametric density in \autoref{eq:barfunc} has a single constant scale height, however the results of \autoref{sec:slicefits} suggest that the scale height is not constant becoming thinner towards the bar end. For this reason we have added a second component parameterised with the same functional form to represent the super-thin component. We construct the luminosity function of the super-thin component assuming a constant star formation rate, as opposed to the 10\Gyr age of the bulge and 180\pc scale height thin bar. As discussed in the previous section this choice reduces the mass-per-clump star compared to a simple 10\Gyr population.

We initially allowed all parameters of the fit to vary, however we found that this resulted in an unsatisfactory fit. The higher number of stars in the fields nearer to the bulge region result in smaller Poisson errors and therefore higher weight placed on fitting these well, so that outside this region the fit is less good. In particular the thin bar component became too thick in order to fit the fields near the bulge, and the resultant vertical structure in plots similar to Figs. \ref{fig:glimpsescaleheightexamples}--\ref{fig:sigmarc} was poorly represented.

\begin{figure*}
\includegraphics[width=\linewidth]{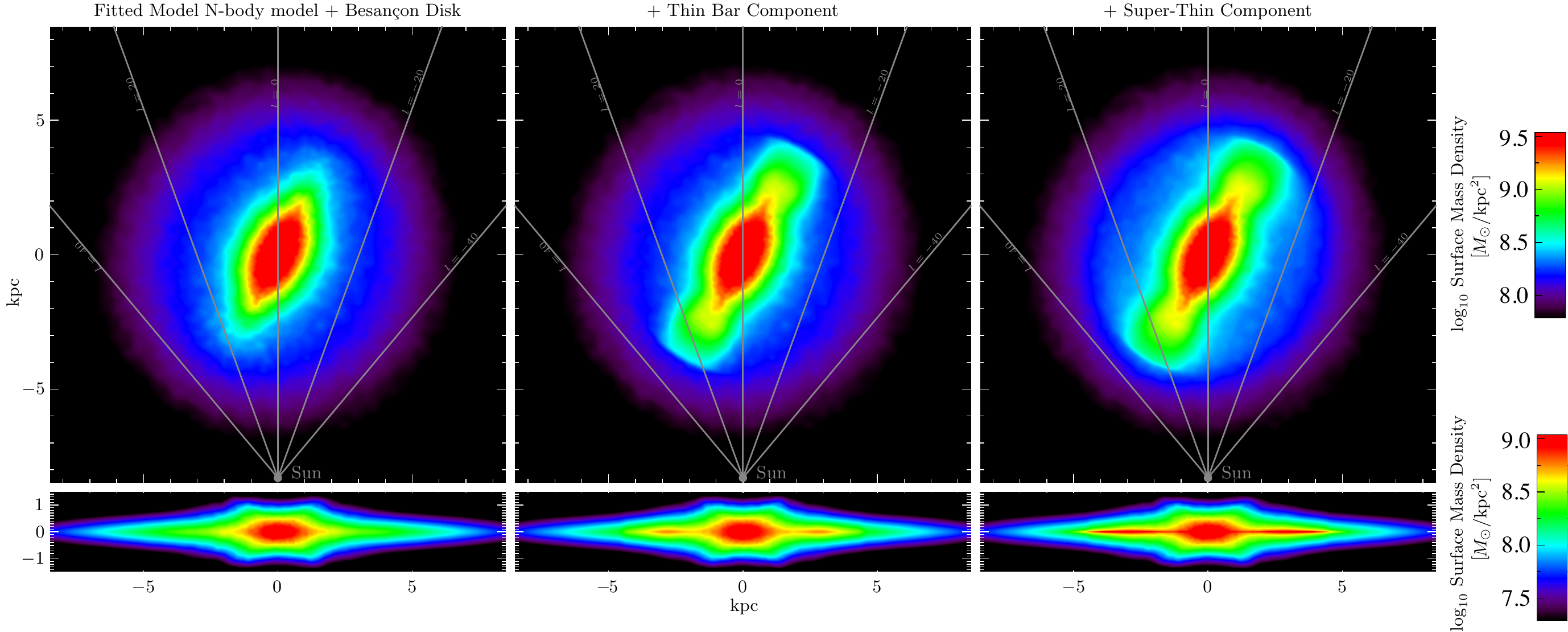}
\caption{The three upper panels show the view from the north Galactic pole of the surface mass density of the best fitting two component model described in \ref{sec:thinfit}. The lower panels show the same model observed side-on. We also add the Besan\c{c}on model disk density described in \citet{Robin:03} since the disk of the N-body model alone is insufficient. The left hand panels show the N-body model together with the Besan\c{c}on model disk. The central panels additionally include the 200\pc thin bar component. The right hand panels additionally include the 40\pc super-thin bar component. The bar length was measured from these images using the methods described in \autoref{sec:barlength}. \label{fig:above}}
\end{figure*}

Instead, motivated by \autoref{fig:expscaleheight} we fixed the vertical scale height of the two components, finding that exponential scale heights of 200\pc and 40\pc produced a reasonable representation of the vertical structure. The resultant fit is considerably improved, particularly near the bar end and between $10\dg<l<20\dg$, as can be seen in the lower panels of \autoref{fig:chi2_frac_all}. We also observe this model similarly to the data, in slices from above, and plot the result as the lower half of \autoref{fig:kaboverho}, finding it does a remarkably good job of reproducing the features in the data. We show in \autoref{fig:above} the fitted two-component model, separated by component,  projected both from above and perpendicular to the bar along the minor axis.

The angles of the component are $\alpha=29.1\dg$ for the thin component and $\alpha=30.0\dg$ for the super-thin component, again consistent with measurements of the bar angle in the bulge region. The mass of the 200\pc thin component is $3.3 \times10^9\msun$ while the mass of the 40\pc super-thin component is $4.0 \times10^9\msun$ giving a total mass together with the N-body bar of $1.81\times 10^{10}\msun$.

\subsection{Bar Angle}

In our fiducial model the fitted bar angle of the parametric long bar ranges from $\alpha=(29-30)\dg$ depending on the number of components (\autoref{tab:parms}). Fitting the other N-body models described in \citet{Portail:15} with the same procedure gives angles in the range $\alpha=(28-31)\dg$.  Altering the age and metallically of the bar components while holding the bulge luminosity function fixed changes the fitted bar angle slightly. This is because the RCG luminosity then changes between the inner galaxy and the long bar region. Making the super-thin component super-solar metallicity ($[z/H]=0.17$) increases the fitted angle of this component to $32.4\dg$, while changing the $200\pc$ component to have a constant star formation history increases its angle to $32.7\dg$. We therefore increase our range of possible long bar angles to $\alpha=(28-33)\dg$ to encompass this.
 This is consistent with alignment with the angle found in the central $|l|<10\dg$ in \citet{Wegg:13} of $\alpha=(27\pm 2)\dg$ which is our assumed N-body bulge angle, and with $\alpha=(29-32)\dg$ found by \citet{Cao:13} for a simpler parametric \citet{Dwek:95} bulge model.

We regard the one component long bar model described in \autoref{sec:barfit} as a useful confirmation that the fitted bar angle is insensitive to the parametric model. This is a significantly simpler model that fits the data less well than the two component long bar described in \autoref{sec:thinfit} but still recovers a consistent bar angle. 

We have performed an MCMC to estimate errors on both the model parameters, and the physical properties such as bar mass and length. The resultant statistical errors are extremely small, significantly smaller than the difference between fits with different models. We therefore consider the statistical errors to be negligible in comparison to systematic errors quoted above. We have also refitted starting from different initial conditions to verify that the fitted parameters such as the bar angle is not a local minimum, or strongly dependent on initialisation position. While the same minimum in $\chi^2$ was not always reached by the minimisation procedure the difference was smaller than between models.

\subsection{Bar Length}
\label{sec:barlength}

In external galaxies and N-body simulations the bar length is dependent on definition with many possible \citep[\eg][]{Athanassoula:02}. Even in N-body simulations with strong bars, where the full 6-dimensional phase space of every particle as a function of time is available with negligible error, the different definitions produce different bar lengths, typically at the level of $\approx 10\%$ for strong bars. We therefore do not expect an unambiguous and definitive bar length for the Milky Way, particularly with the more challenging viewing geometry. Instead we estimate the bar length of our best fitting model densities using several definitions. These bar length measurements are summarised in \autoref{tab:physicalparms}.

Specifically we use three bar length definitions found in \citet{Athanassoula:02} to give reasonable bar lengths of N-body models:
\begin{inparaenum} 
\item $L_{\rm drop}$, the point where the ellipticity of the face on profile drops fastest.
\item  $L_{m=2}$, the point where the $m=2$ Fourier mode of the face on image drops to 20\% of its maximum.
\item $L_{\rm prof}$ the point where the major and minor axis profiles agree within 30\% \citep[this a larger threshold than the 5\% used by][which we found gave spurious results for our densities]{Athanassoula:02}.
\end{inparaenum}

We add one additional bar length measurement, $L_{\rm mod}$. We take the difference between the face-on major and minor axis surface density profiles along the bar and fit an exponential to the long bar region. We then define the bar length as the point at which the density falls to $1/e$ of the exponential profile. In the case of an analytic exponential bar with a Gaussian cutoff, like our parametric long bar functions, this corresponds to defining the bar length as $R_o+\sigma_o$.
  
All these methods were applied to face-on images of our densities to which we added the density of the disk in the Besan\c con model. We show these face-on images in \autoref{fig:above} and the resultant bar half lengths are given in \autoref{tab:physicalparms}. All our stated bar lengths are the half length, defined as the distance from the galactic centre to the bar end.

\begin{figure*}
\includegraphics[width=0.75\linewidth]{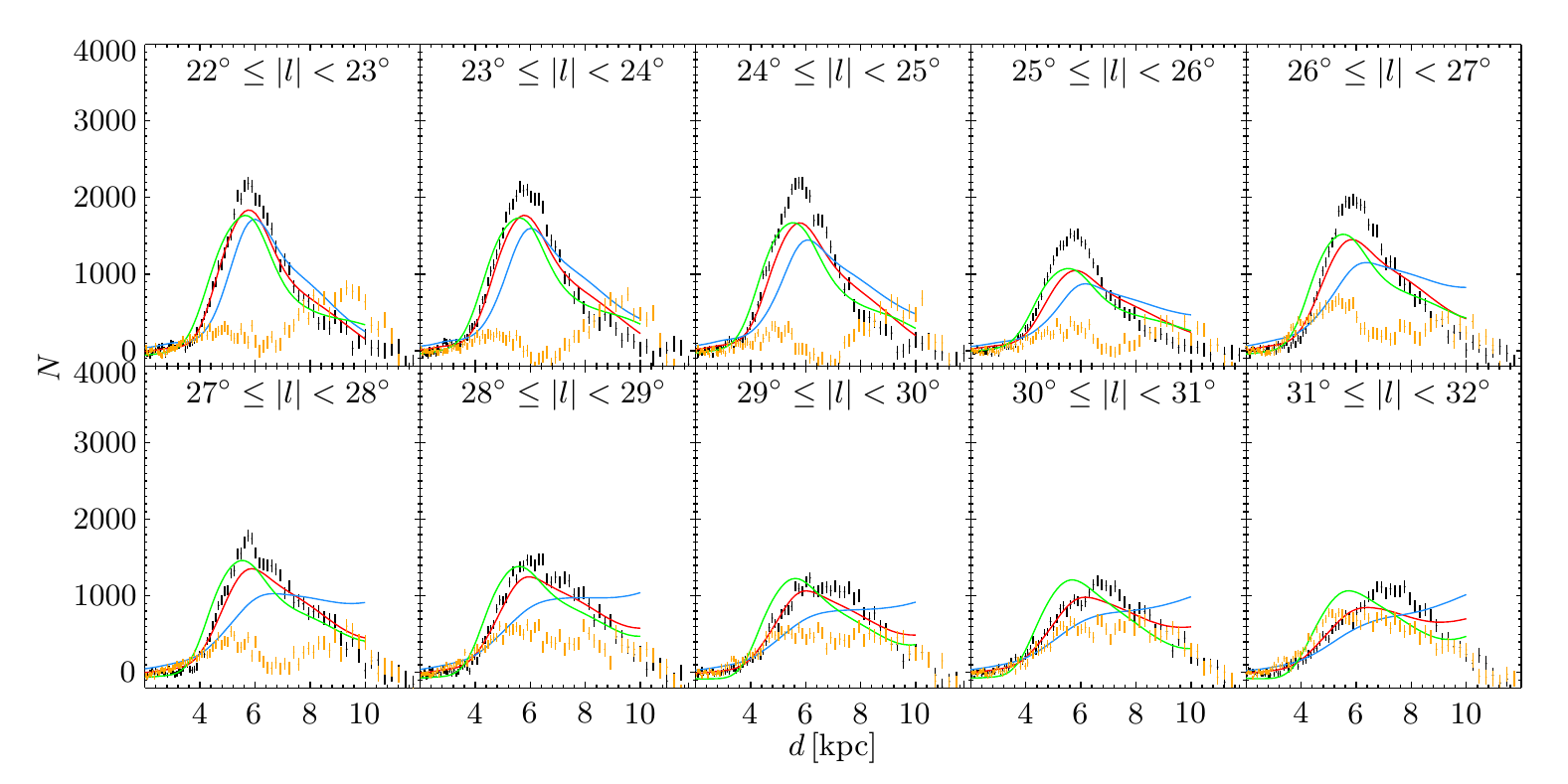}
\caption{\label{fig:barenddiag}Histograms of stars near the bar end. In black we show histograms of stars with positive longitudes and $0.15\dg \le |b| < 1.35\dg$ in one degree longitude intervals. The histograms of $\mu_K$ (\autoref{eq:mujhk}) were converted to distance assuming all stars are red clump stars and an exponential in $\mu_K$ was subtracted for visualisation, which can be thought of as representing non-red clump stars. In orange we show the same procedure applied to the data at negative longitues to highlight the non-axisymmetric features. In red we show our best fitting  density model from \autoref{sec:fitrho}, while in blue and green we show the same model but with the bar length reduced and increased by 0.5\kpc respectively.  }
\end{figure*}

Before considering the bar length measurement we first return to the data near the bar end. In \autoref{fig:barenddiag} we show histograms stacked in Galactic latitude as a function of longitude. We show both the positive and negative longitude sides to demonstrate the peak at positive longitudes is non-axisymmetric. To make the plot clearer we also subtract an exponential in $\mu_K$ which can be thought of as representing the background of non-RCGs. The bar is clear and well localised to $l<26\dg$. At $l>30\dg$ while non-axisymmetry still appears it is much less significant and fainter than would be expected for the bar. In the region $26\dg<l<30\dg$ the non-axisymmetric excess weakens, broadens and becomes fainter. We therefore presume that the bar ends in this region, possibly transitioning into the spiral arms. If we convert these longitudes of the bar end to a bar length assuming that the bar lies at $\alpha\approx 27\dg$ and that projection effects are negligible we would recover a bar half length between 4.4 and 4.8\kpc. 

We  produced a similar plot to \autoref{fig:barenddiag} showing just the N-body model and it is clear that the bar ceases to be significant at too low longitude. In contrast plotting the one component long bar model it is clear that the bar extends beyond the data in longitude to where there is no non-axisymmetry in the data. For this reason we disregard the bar length measurements of these models in \autoref{tab:physicalparms}.

Instead the two component model is a significantly better fit to the stacked data near the bar end in \autoref{fig:barenddiag}. We show it compared to the data together with variations in which the bar was artificially lengthened and shorted by adjusting the outer cutoff by 0.5\kpc. The model with an artificially shortened bar is insufficient particularly beyond $l=25\dg$. In contrast the model with an artificially lengthened bar predicts excessive non-axisymmetry beyond $l=30\dg$ when the positive and negative latitudes have similar counts at the distance of the bar.

Because the two component model appears to reasonably fit the stacked data in \autoref{fig:barenddiag} we consider the measurements of this model to be our fiducial bar half length. These measurements lie in the range $4.73-5.23\kpc$. We have repeated this process on the other N-body models finding that the variation between models is smaller than the variation between methods of measuring bar length. The one component bar length appears longer, however it is evident from \autoref{fig:chi2_frac_all} that this model fits poorly in the region beyond $l>30\dg$. Therefore taking the average and standard deviation of these measurements we consider our fiducial estimated bar half length for the Milky Way to be $(5.0 \pm 0.2)\kpc$.

\section{Discussion}
\label{sec:discuss}

\subsection{Continuity Between Box/Peanut Bulge and Long Bar}

Two lines of evidence in this work support that bar and bulge appear to be naturally connected: the angle between the Box/Peanut (B/P) Bulge and Long Bar is small, and the scale height along the bar decreases smoothly. 

We find in this work that the long bar has bar angle in the range $\alpha =  (28-33)\dg$ consistent with recent determinations of angle found at $|l|<10\dg$ in the B/P Bulge \citep[\eg][]{Wegg:13}. We find this angle by fitting the magnitude distributions through the parametric density functions in \autoref{sec:fitrho}. It was already clear however directly from the data in \autoref{fig:kaboverho} that the bar angle between $10\dg<l<20\dg$ was close to the $\alpha\approx 27\dg$ found in the B/P Bulge.

The alignment of the B/P Bulge and long bar in this work is in contrast to some previous claims of misalignment \citep{LopezCorredoira:06,CabreraLavers:07,CabreraLavers:08}. 
\citet{Benjamin:05} found a large angle of the long bar of $\alpha=(44\pm10)\dg$ using GLIMPSE data. However the GLIMPSE long bar data was subsequently analysed finding $\alpha=(38\pm 6)\dg$ (\citealt{Zasowski:thesis}, or $\alpha=35\dg$, \citealt{Zasowski:12}) due to a fainter assumed RCG absolute magnitude. This results in the derived distances to the long bar being reduced and therefore the bar angle decreasing for the same Sun--Galactic center distance, $R_0$. Note that because we use data that extends in longitude to the bulge region and allow the red clump absolute magnitude to vary in our fitting we are insensitive to this degeneracy in this work.
Some misalignment claims appear to be a result of using the endpoints to derive the bar angle \citep[\eg][]{GonzalezFernandez:12,deAmores:13}, however the projection effects can be extreme especially for the far end of the bar. Large bar angles from fitting the distance of clump giants \citep[\eg][]{CabreraLavers:07,CabreraLavers:08} are more difficult to reconcile with this work and may be related to the very different selection criteria. In those works RCGs are selected from the colour-magnitude diagram together with an extinction model which predicts the reddening as a function of distance modulus, while in this work they are statistically identified as an excess above the smooth background of non-RCGs.

\begin{figure}
\includegraphics[width=\linewidth]{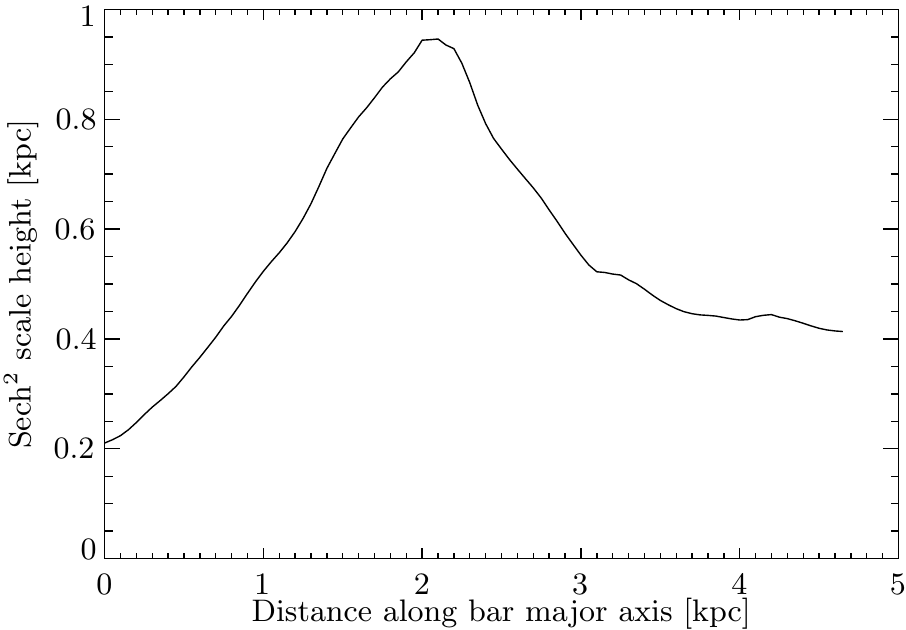}
\caption{The scale height of the N-body model M85 from \citet{Portail:15} as a function of distance along the major axis of the bar. We use a ${\rm sech}^2$ scale height since this is a better fit to the N-body vertical structure than an exponential. The model shown is  prior to fitting the bulge data to demonstrate that the transition from short central scale height, to large scale height in the B/P region, to short scale height in the long bar region arises naturally. \label{fig:modelscaleheight}}
\end{figure}

In addition we find that the scale height shown in \autoref{fig:expscaleheight} appears to smoothly transition between the B/P bulge and the long bar which also suggests that both are part of the same connected structure. This transition is similar to N-body models. To demonstrate this we show in \autoref{fig:modelscaleheight} the scale height of one of the initial barred N-body models. The transition in vertical structure from a short central scale height, to a large scale height through the 3D B/P region, to a short scale height in the long bar can be seen to arise naturally in the N-body model prior to fitting the bulge data.

Given the near alignment it seems natural that the long bar is the in-plane extension of a vertically extended inner part of the bar, the Box/Peanut bulge. This is similar to the structures found in N-body simulations where the buckled three-dimensional boxy/peanut bulge is shorter than the bar \citep[\eg][]{Athanassoula:05,Inma:06}, as well as in external galaxies \citep{Erwin:13}. This situation was previously suggested in the Milky Way by \citet{MartinezValpuesta:11} and \citet{Romero:11} despite the possible misalignment.

In the lower panel of \autoref{fig:above} we show the side-on projection of the best fitting model. These side projections where the three dimensional peanut has a lesser extent than the bar is similar to the side projections in \citet{Athanassoula:05} and the observations of \citet{Bureau:06}. 

\subsection{Thin and Super-thin Bar Component}

From fits to the vertical profile of the RCGs we find evidence for two components associated with the long bar of the Milky Way. A component with a scale height of $\approx 180\pc$ which we term the `thin bar' since it would appear to be formed from the counterpart of old solar neighbourhood stars, and a second `super-thin bar' component with scale height $\approx 45\pc$.

The density better fitting a broken exponential alone would be weak evidence for the presence of two distinct components: As demonstrated by the controversial thin/thick disk discussion in the solar neighbourhood it is possible for a continuum stellar distribution to mimic a double exponential in vertical profile \citep[\eg][]{Bovy:12}. However the relative contributions of the thin and super-thin components change along the bar. The thin component decreases outwards approximately exponentially with a scale length similar to the Milky Way disk, while the super-thin component \emph{increases} outwards towards the bar end. This spacial mismatch strengthens the argument for two distinct components. Additional abundance and age measurements would strengthen the argument further.
 
The short scale height for the super-thin bar is similar to the 60-80\pc found in \Kband imaging of NGC 891 by \citet{SchechtmanRook:13}. It is slightly lower than the expected value in the solar neighbourhood. Young stars locally have a scale height of $\sim 60\pc$ \citep[\eg][from measurements of OB stars and young open clusters]{Joshi:07} and the vertical disk heating in the last 2\Gyr is insignificant \citet{Holmberg:2007}.

If we extrapolate the local thin disk properties inwards we can predict the resultant velocity dispersion of the components. Assuming that the thin bar's scale height is set by its self gravity then the vertical dispersion, $\sigma_z$, will vary as $\sigma_z^2 \propto \Sigma h$ where $\Sigma$ is the surface density and $h$ the scale height. Assuming that the surface density $\Sigma$ is exponential $\Sigma \propto \exp \left(R/R_d\right)$ then the surface density at Galactocentric radius 4\kpc is a factor $7.4$ higher than in the solar neighbourhood for $R_d=2.15\kpc$ \citep{Bovy:13}. Since the local thin disk scale height (\eg 225\pc: \citealt{Veltz:2008}, 300\pc \citealt{Juric:08}) is slightly larger than the thin bar scale height the resultant dispersion would then be a factor $\sim \sqrt{7.4}\times 180/225=2.2$ larger, or approximately $\sigma_z\approx 46 \kms$ assuming a local vertical velocity dispersion of $21\kms$ \citep{Binney:14}.

Estimating the velocity dispersion of the super-thin component is less certain since it depends on the extent to which its vertical structure is governed by self-gravity. We consider the two extremes of completely self-gravitating and non-self-gravitating and expect these to bracket the true structure. Assuming that it is self-gravitating a similar extrapolation gives $\sigma_z \approx 30\kms$, assuming locally that stars younger than 2\Gyr have $\sigma_z\approx10\kms$ \citep{Holmberg:2007} and a scale height $\sim 60\pc$ \citep{Joshi:07}. The dispersion of a non-self-gravitating structure  would be lower. Assuming that the potential is near harmonic $\Phi \propto z^2$ and that the vertical frequency is set by the thin component then the vertical frequency will be $\nu_z \propto \sigma_z/h$ and hence 2.7 times higher than in the solar neighbourhood. The resultant dispersion for the super-thin component would be $\sigma_z \propto \nu_z h$ and therefore $\sigma_z \sim 2.7 \times 50\pc/60\pc \times 10\kms \approx 22\kms$. 

The origin of the super-thin component is unclear. The small scale height means the stars cannot have experienced much vertical heating and therefore should be younger than the thin component. To have formed RCGs the stars must still be $\gtrsim 500\Myr$ old but galaxies with ongoing star formation have a strong bias to younger RCGs with a peak of age $\sim 1\Gyr$ \citep{Girardi:01,Salaris:02}. They could  be related to star formation towards the bar end \citep{Phillips:96}. Because their lifetime is longer than the orbital timescale the interpretation would be that they are stars on bar following orbits that formed towards the bar end and spend more time towards their apocenter near the bar end. Even without any star formation, since the bar grows with time \citep[\eg][]{Inma:06}, the bar will have grown into the star forming disk. These recently captured bar stars would then spend more of their orbital period at their apocenter near the bar end and could therefore have a similar density profile to the super-thin component. Further dynamical and chemical information is needed to distinguish between these possible scenarios.

\subsection{Bar Length}
\label{sec:barlengthdiscuss}

Our estimated bar half length of $(5.0 \pm 0.2)\kpc$ is larger than most previous estimates \citep[\eg 4.5\kpc by][]{CabreraLavers:08}. This is partly because our bar angle is less than most previous estimates. If one assumes that the super-thin component is not part of the bar the same bar length estimators applied only to the thin component would lead to an only slightly shorter bar length of $(4.6 \pm 0.3)\kpc$. Without dynamical information the exact nature of this region near the bar end and the possible transition to spiral arms or ring structures is difficult to determine.

A further uncertainty is that in estimating bar lengths we use the face on view of our model together with the Besan\c con model disk. If the Besan\c{c}on disk density is not accurate in this region it would slightly impact our estimated bar length. For example increasing the density by a factor of 2 reduces the bar length from $(5.0 \pm 0.2)\kpc$ to $(4.6 \pm 0.2)\kpc$. This is why $L_{mod}$, the point where the difference between the major and minor axis profiles of the face on surface density drops to $1/e$ of the otherwise smooth and slower decline along the bar, is useful. Since it considers only the non-axisymmetric part of the density it is independent of details of the disk.

A firm lower limit to the bar length can be determined visually  from \autoref{fig:kaboverho}: it is clear from the slices at $b>2\dg$ that the bar extends to $l\gtrsim 22\dg$ and therefore given a bar angle $\alpha=30\dg$ the bar cannot be shorter than $\approx 4.0\kpc$.

\begin{figure}
\includegraphics[width=\linewidth]{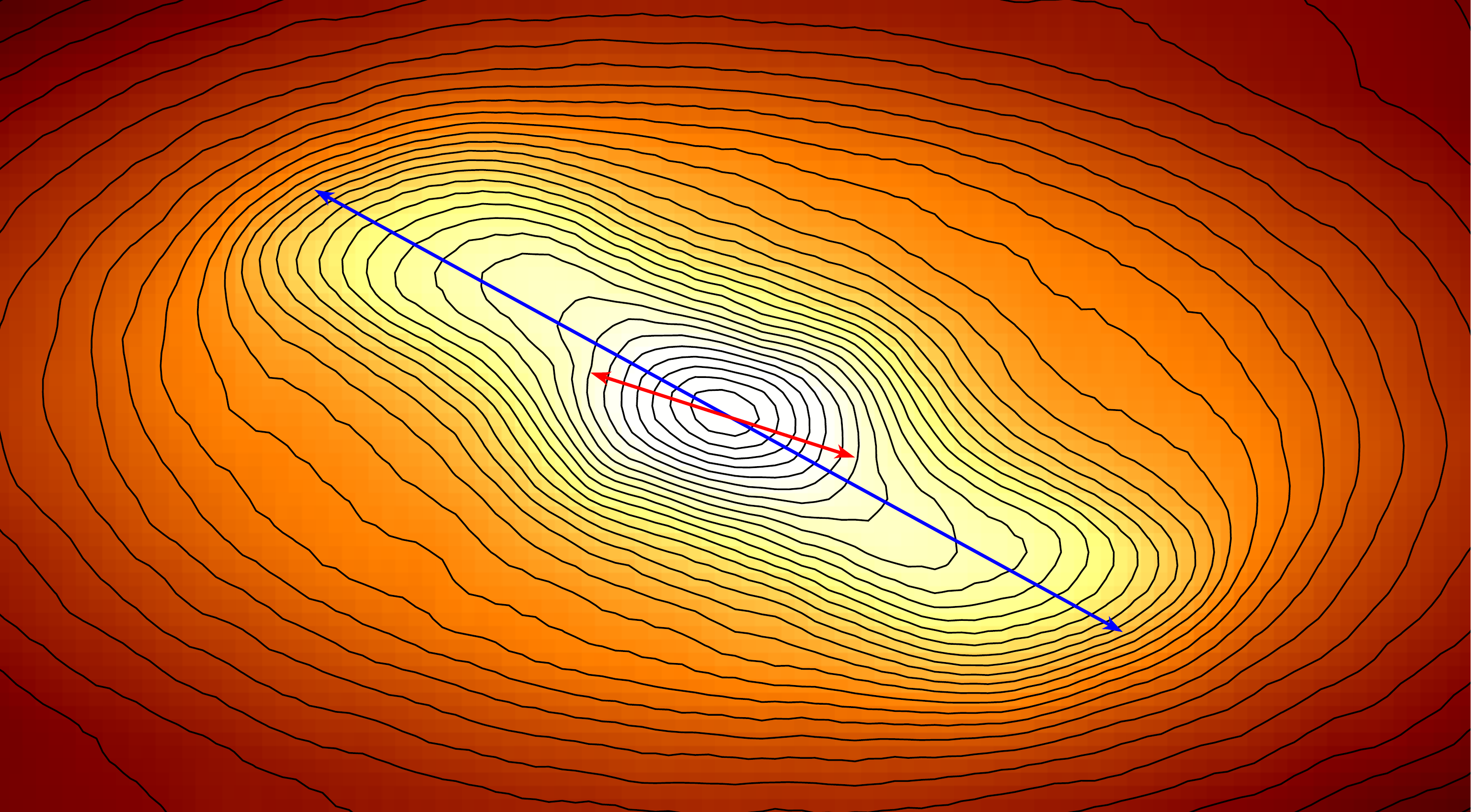}
\caption{Our fitted two component Milky Way model density (\autoref{sec:thinfit}) observed as if an external galaxy with bar angle 45\dg and inclined at $i=60\dg$. The line-of-nodes of the disk is horizontal. As described in \citet{Erwin:13}, this is an orientation where both the boxy and barred zones are visible with the angle of the boxy zone closer to the line-of-nodes than the projected angle of the long bar. The full length of the bar and boxy zone were measured from this image visually in a manner similar to \citet{Erwin:13} and are marked in blue and red respectively. \label{fig:erwin}}
\end{figure}

External galaxies show a strong correlation between the size of their boxy bulge, and their in-plane long bar which is seen as `spurs' extending from the boxy region. \citet{Erwin:13} finds the ratio of the size of the boxy 3D bulge to the thinner in-plane bar is $0.42\pm 0.07$. This was estimated by measuring a sample of moderately inclined external galaxies. We therefore inclined our model density and measured the boxy zone and spurs in a similar manner. In \autoref{fig:erwin} we place an external observer of the Milky Way inclined at $60\dg$ with the bar at $\Delta {\rm PA}=45\dg$ to their line of sight. We find they would measure a box bulge size of $2 R_{\rm box}=2.3\kpc$, a bar length $2L_{\rm bar}=8.8\kpc$ and therefore a ratio to the bar length of $0.26$. This is towards the lower end of the external galaxies measured by \citet{Erwin:13}, at the edge of the observed range.

The bar length measured in this work here has implications for the pattern speed of the bar. The dimensionless bar rotation velocity is $\mathcal{R}\equiv\frac{R_{\rm CR}}{L_{\rm bar}}$ where $R_{\rm CR}$ is the corotation radius. Since the bar cannot extend beyond corotation $\mathcal{R} \geq 1$ and our measurement of bar length implies corotation must lie outside $(5.0 \pm 0.2)\kpc$. In turn this limits the pattern speed to be lower than $\lesssim (45 \pm 2)\, {\rm km}\,{\rm s}^{-1} \kpc^{-1}$ for a flat rotation curve near the bar end with circular velocity $v_c=218\kms$ \citep{Bovy:12b}. This is in some tension with the pattern speed derived form the interpretation of the Hercules stream as being due to the outer Linblad resonance of the bar. \citet{Antoja:14} find a pattern speed of $(48.2\pm0.5)\kms$ if the bar lies at $\alpha=29\dg$ with $R_0=8.3\kpc$ and circular velocity in the solar neighbourhood of $v_c=220\kms$. Reconciling these results requires either an extremely fast bar with $\mathcal{R}\approx 1$, where the bar ends at corotation, or that the Hercules stream has a different origin than the outer Lindblad resonance. In contrast, by constructing dynamical models of the B/P bulge, \citet{Portail:15} found $25-30\, {\rm km}\,{\rm s}^{-1} \kpc^{-1}$. Combining this with our bar length and its error gives $\mathcal{R}$ in the range $1.4-1.9$ and therefore the Milky Way would be a slow rotator \citep[$\mathcal{R}>1.4$,][]{Rautiainen:08}.

In the future we plan to combine the constraints from this work with kinematic data in the long bar region and perform made-to-measure modelling in a manner similar to \citet{Portail:15}. Constructing self-consistent models of the long bar with added kinematic data will better indicate which stars are on bar following orbits, and thus constrain the dynamical structure near the bar end and the bar length.

In addition the Gaia mission \citep{deBruijne:12} is capable of some elucidation of the long bar. Extinction will limit the effectiveness of its broadband visual measurements near the galactic plane. However the result that the long bar extends out of the plane means that at $|b|\gtrsim 1\dg$ it should be possible to measure reliable parallaxes and proper motions for a sufficient number of bright RGB stars.  In addition GAIA will allow for better characterisation of RCGs as standard candles, and therefore further their use in studying Galactic structure in the higher extinction regions, where NIR measurements are required. In particular, better characterisation of the population effects on RCGs will be of importance in the future since, for example, our largest uncertainty in bar angle is due to the uncertainty in the difference in RCG magnitude between bar components.

\section{Conclusion}
\label{sec:conclude}

We have investigated the structure of the Milky Way's bar outside the bulge, termed the long bar, using red clump stars as a standard candle and tracer of the underlying density. We have combined data from UKIDSS, VVV, 2MASS and GLIMPSE and have constructed magnitude distributions in many fields covering the central $|l| <40\dg$, $|b|<9\dg$. We concentrated on the \Kband and corrected for extinction star-by-star. On the basis both of fitting the clump stars  in each field individually, and of constructing parametric density models that simultaneously fit all fields we reach the following key conclusions:
\begin{itemize}
	\item The bar extends to $l \sim 25 \dg$ at $|b| \sim 5\dg$ from the Galactic plane, and to $l \sim  30 \dg$ at lower latitudes. 
	\item The long bar of the Milky Way lies at an angle of $\alpha =  (28-33)\dg$. This is consistent with being aligned to the Milky Way's Bulge \citep[\eg][]{Wegg:13,Cao:13}. We find this angle from the parametric density fits, but it is also visually evident from the raw data in \autoref{fig:kaboverho}. 
	\item The overall scale height of the bulge transitions from a short central scale height, to large scale height in the B/P region, to short scale height in the long bar region. This transition arrises naturally in N-body models of barred galaxies, and together with the alignment of the long bar and B/P region indicates that the bar and bulge are part of a single innately connected structure. 
	\item We find evidence for two scale height components. We find a $\approx 180\pc$ `thin bar' component, which decreases in density outwards, and which we interpret as the barred counterpart of the solar neighbourhood thin disk. In addition there is a $\approx 45\pc$ `super-thin bar' component whose density increases outwards along the bar. This component is likely to be related to younger stars ($\sim 1\Gyr$) towards the bar ends. 
	\item The offset in $b$ of the bar is consistent with symmetry about the physical Galactic mid-plane under the assumption that the sun lies 25\pc above the Galactic plane and \sgra lies in the physical Galactic plane. The vertical structure is also consistent with the bar lying in the physical Galactic mid-plane to $\lesssim 5\pc$.
	\item We find a bar half length of $(5.0 \pm 0.2)\kpc$ measured by applying commonly used bar length definitions to our best fitting parametric model. This agrees with simple estimates directly from the data near the bar end (\autoref{fig:barenddiag}). The bar length is still somewhat ambiguous since it depends on the interpretation of the nature of the super-thin bar, and excluding this results in a bar length of $(4.6 \pm 0.3)\kpc$. It is also visually clear from the data in \autoref{fig:kaboverho} that the bar appears straight extending to at least  $l\approx 22\dg$ and therefore cannot be shorter than $\approx  4.0\kpc$ as a lower limit. 
	\item Our estimated total bar mass in our fiducial parametric model is $1.8\times 10^{10}\msun$.  This arises from the number of clump stars in the bar together with a `mass-to-clump' ratio arising from our assumed IMF and component ages. Varying these assumptions dominates the error. For example changing from our assumed \citet{Kroupa:01} IMF to a \citet{salpeter:55} IMF increases the long bar mass by a factor 1.43, while changing from an old 10\Gyr stellar population to a constant star formation rate reduces the mass by a factor of 2.0.
\end{itemize}
\FloatBarrier

\section{Acknowledgments}

We are grateful to Inma Martinez-Valpuesta for providing initial N-body models and we acknowledge useful discussions with Bob Benjamin and especially Matias Bla\~{n}a.

\bsp

\appendix
\section{Transformations to 2MASS photometric system}
\label{sec:xforms}

\subsection{Transformation from UKIDSS}

The UKIDSS survey is performed in the WFCAM photometric system. From DR2 onwards the zero points were calibrated from 2MASS but the photometric system is slightly different \citep[equations 4-8 of][]{Hodgkin:09}. The most straightforward method of transforming WFCAM magnitudes to 2MASS would be to use the inverse of the calibration transformations. However these rely on the $J$-band magnitude and the $J$-band would then be the limiting band in high extinction regions, having almost twice as much extinction as the $H$-band. Instead for the $H$ and $K$ bands we measure the transformation from the $H-K$ color for giants, which can then be applied directly in equation \ref{eq:mujhk}. 

To derive the transformation we cross match 2MASS and UKIDSS sources within 1\arcsec and 1\mags with 2MASS photometric quality classified as `A' or `B' in the $J$, $H$ and $K$ bands. A cut was applied in color-color space of UKIDSS to remove nearby dwarfs and select mostly giants: $0.25<(J-K)-2.5(H-K)<0.55$. In order to avoid saturation in UKIDSS and Eddington bias (over representation of the more numerous faint objects due to measurement error) in 2MASS we also select only sources with $11<K<13.5$,  $12.5<H<15$ and $12<J<15.5$. 

\begin{figure}
\includegraphics[width=80mm]{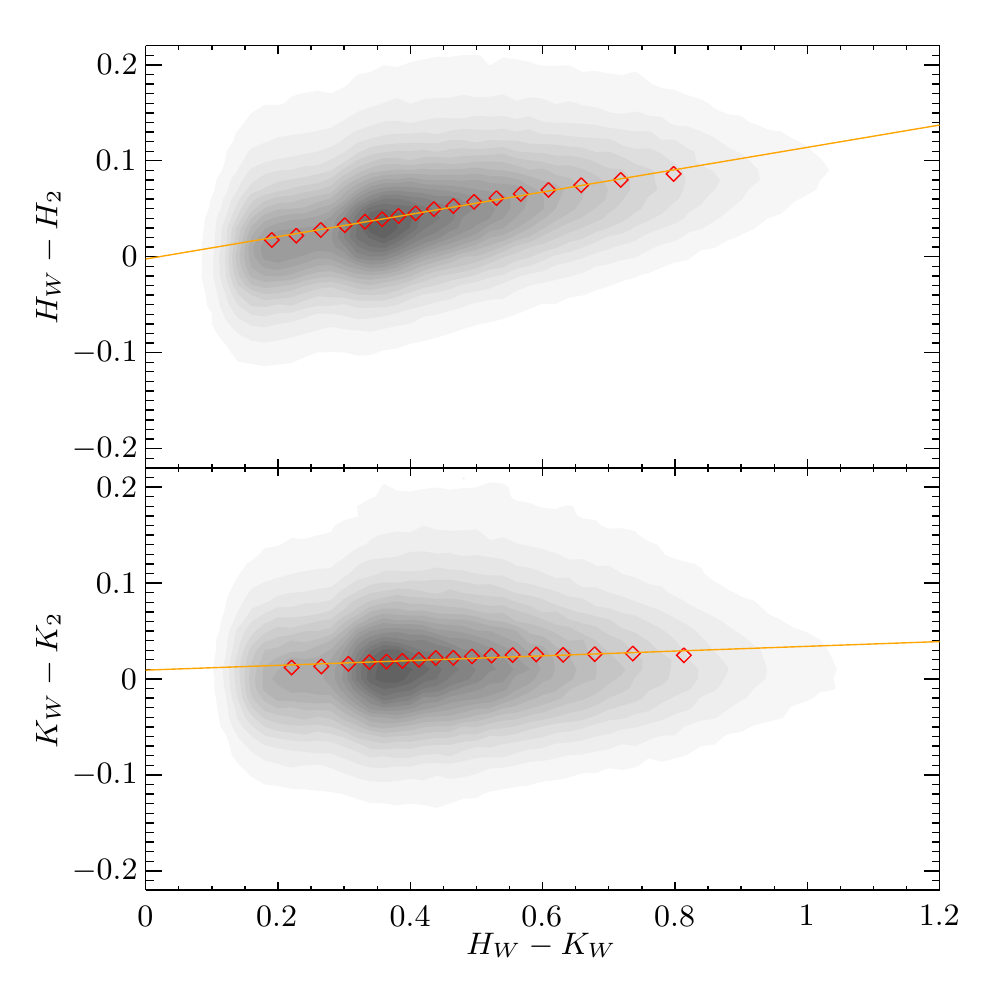}
\caption{Comparison of photometry from 2MASS (subscript $2$) and UKIDSS (subscript $W$) for the $K$ and $H$-bands. Contours show the density of cross matched sources selected as described in the text. Red points are the binned sigma clipped mean and the orange line is the derived photometric transformation.  \label{fig:hxform}}
\end{figure}

\begin{figure}
\includegraphics[width=80mm]{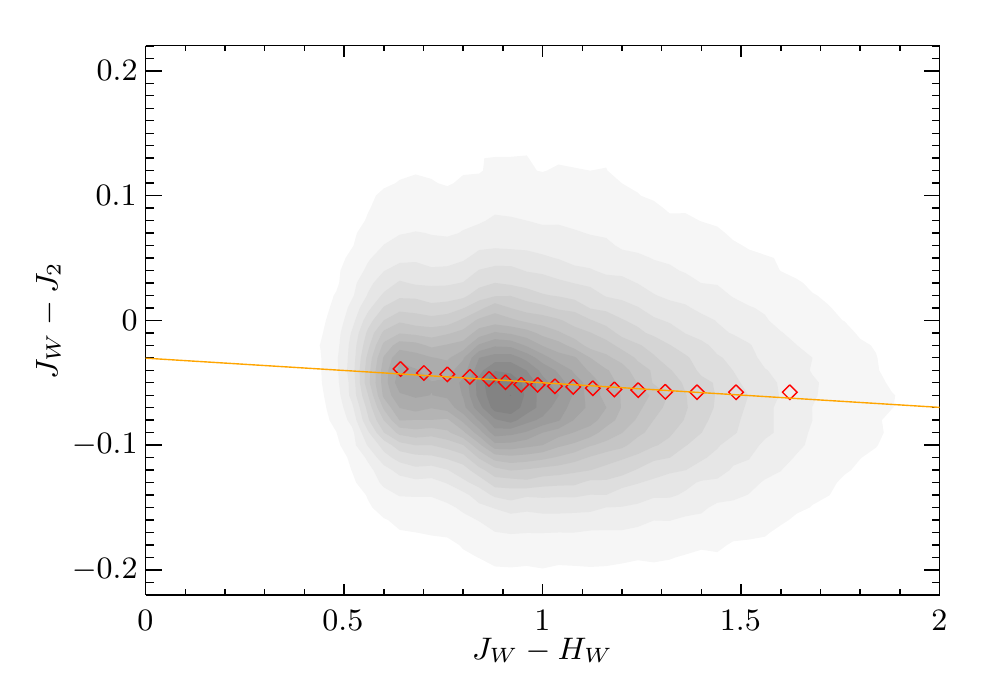}
\caption{Comparison of photometry from 2MASS (subscript $2$) and UKIDSS (subscript $W$) for the $J$-band. Contours show the density of cross matched sources selected as described in the text. Red points are the binned sigma clipped mean and the orange line is the derived photometric transformation. \label{fig:jxform}}
\end{figure}

With this sample of cross-matched sources we then derive the photometric transformations. We divide the sample into 20  color bins each with equal numbers of stars and in each calculate the difference in magnitude between 2MASS and UKIDSS using an iterative sigma-clipped mean at the 2.5 sigma level. Our derived transformation is then the linear regression of these points, neglecting the first and last bins to reduce outliers. This process is shown applied to the $H$ and $K$ bands in figure \ref{fig:hxform}, and the $J$ band in \ref{fig:jxform}.

The resultant transformations are:
\begin{align}
J_2 &=J_W - 0.02(J_W - H_W) - 0.03 \nonumber \\
H_2 &=H_W + 0.12(H_W - K_W) \\
K_2 &=K_W + 0.02(H_W - K_W) + 0.01 \nonumber
\end{align}
where subscript $2$ refers to the 2MASS system and $W$ the WFCAM system used in UKIDSS.

\subsection{VVV Transformation}

The VVV survey is performed in the VISTA/VIRCAM photometric system. As for the WFCAM system this is tied to, but different from the 2MASS system\footnote{For details see \url{http://casu.ast.cam.ac.uk/surveys-projects/vista/technical/photometric-properties}.}. In principle the same method as for the WFCAM system could also be used for VISTA. However the VVV system is significantly closer to the 2MASS system making this less necessary. There is however a field-to-field scatter in high extinction regions where there can be variations in zero point of $\sim 0.1\mags$. We instead take a different approach and as in \citet{gonzalez:11b} and \citet{Wegg:13} we re-estimate the zero points for each field by cross matching bright but unsaturated VVV stars with 2MASS.

\section{GLIMPSE data}
\label{glimpseappend}

\begin{figure*}
\includegraphics[width=0.9\linewidth]{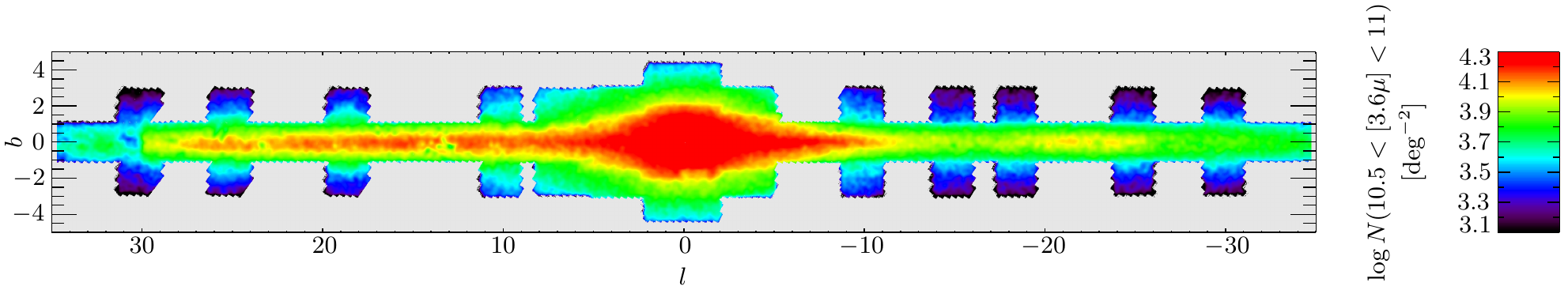}
\caption{The surface density of stars in the GLIMPSE $[3.6\mu]$-band over the extinction corrected magnitude range $10.5 < [3.6\mu]_0 < 11$. Asymmetric number counts in $l$ close to the plane demonstrate non-axisymmetry. Extinction is corrected using the RJCE method  \ie the $H-[4.5\mu]$ colour excess as in equation \ref{eq:muglimpse} and data outside the colour bar range are plotted at its limit. Grey areas are those not covered by the GLIMPSE survey. The equivalent plot for the $K$-band is shown in \fig{extcormap}. \label{fig:glimpseextcormap}}
\end{figure*}

\begin{figure*}
\centering
\includegraphics[width=0.8\linewidth]{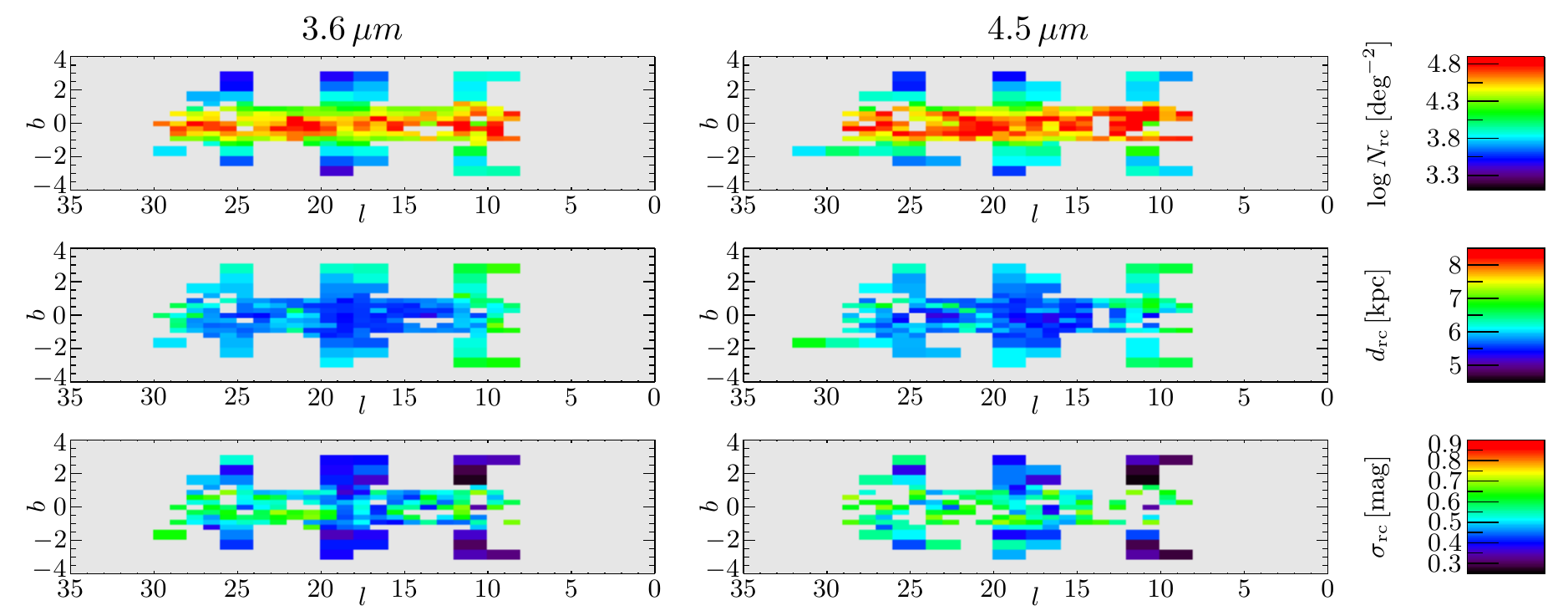}
\caption{As \fig{kfits} but for the GLIMPSE \mone (left hand panels) and \mtwo data (right hand panels). \label{fig:glimpsefits}}
\end{figure*}

\begin{figure*}
\includegraphics[width=0.8\linewidth]{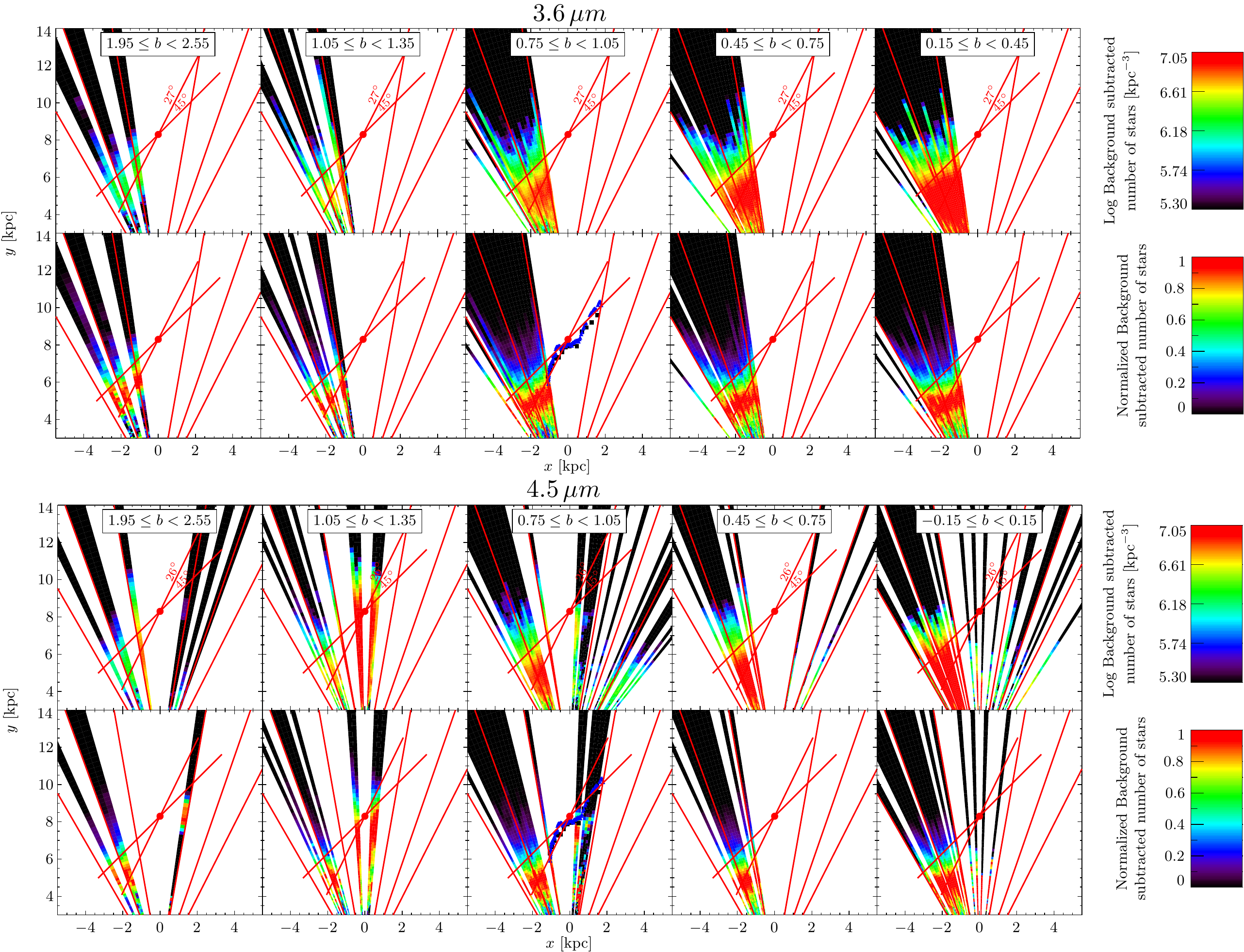}
\caption{\label{fig:3_6aboverho} As \autoref{fig:kaboverho} for the long bar of the Milky Way viewed from above using the GLIMPSE data in the \mone band (above) and \mtwo band (below).}
\end{figure*}

In this section we show the results of GLIMPSE 3.6\micron~and 4.5\micron~data with a similar analysis as applied to the $K$-band in the main text. Although the GLIMPSE data does not have the wide coverage of the $\K$-band data, the reduced extinction in GLIMPSE makes it useful to corroborate the results of the NIR data. In \autoref{fig:glimpseextcormap} we plot the surface density of stars over a narrow range of extinction corrected magnitudes, the equivalent to \autoref{fig:extcormap}. In \autoref{fig:glimpsefits} we show the fitted parameters of the statistically identified red clump stars: their number density $N_{\rm rc}$, the distance $d_{\rm rc}$ calculated from $\mu_{\rm rc}$, and their dispersion $\sigma_{\rm rc}$. This figure is the equivalent to \autoref{fig:kfits} in the \Kband. Completeness limits the reliable identification of RCG stars to $l\gtrsim 8\dg$.  In \autoref{fig:3_6aboverho} we show the equivalent top down view of the raw data as was made for the \Kband in \autoref{fig:kaboverho}. As for the \Kband a bar angle of $\alpha=27\dg$ appears close to the data while $45\dg$ does not. Near the dispersion of the fitted RCG is  larger than the \Kband. This is expected because the RCG fits lie close to the magnitude limit of GLIMPSE in this more crowded and higher extinction area. 

\label{lastpage}
\end{document}